\documentclass[useAMS,usenatbib]{mn2e}
\usepackage{epsfig}
\usepackage{lscape}
\usepackage{epsfig}
\usepackage{rotating}
\usepackage{setspace}

\title[kinematic of low-mass MG candidates]{Spectroscopy and kinematics of low-mass members of young moving groups}

\author[M. C. G\'alvez-Ortiz et al.]{M. C. G\'alvez-Ortiz,$^{1}$\thanks{E-mail:M.Galvez-Ortiz@herts.ac.uk} J. R. A. Clarke,$^{1}$ D. J. Pinfield,$^{1}$ 
 J. S. Jenkins,$^{2}$, S. L. Folkes,$^{1}$ 
\newauthor  A. E. Garc\'ia P\'erez,$^{3}$ A. C. Day-Jones,$^{2}$ B. Burningham,$^{1}$  H. R. A. Jones,$^{1}$  
  J.R. Barnes,$^{1}$ \newauthor and R.S. Pokorny$^{4}$ \\
$^{1}$ Centre for Astrophysics Research, Science and Technology Research Institute, University of Hertfordshire, Hatfield AL10 9AB, UK\\
$^{2}$ Department of Astronomy, Universidad de Chile, Casilla Postal 36D, Santiago, Chile\\
$^{3}$ Department of Astronomy, University of Virginia, P.O. Box 400325, Charlottesville, VA 22904-4325, USA\\
$^{4}$ Yunnan Observatory, P.O. Box 110. CAS. 650011 Kunming. P. R. China.}
\begin{document}

\date{Accepted. Received ; }

\pagerange{\pageref{firstpage}--\pageref{lastpage}} \pubyear{2009}

\maketitle

\label{firstpage}

\begin{abstract}
   We study a target sample of 68 low-mass objects (with spectral types in the range M4.5-L1) previously selected
 via photometric and astrometric criteria, as possible members of five young
 moving groups: the Local Association (Pleiades moving group, age=20 - 150 Myr),
the  Ursa Mayor group (Sirius supercluster, age=300 Myr), the Hyades
  supercluster (age=600 Myr), IC 2391 supercluster (age=35 - 55 Myr) and the 
 Castor moving group (age=200 Myr).
In this paper we assess their membership
 by using different kinematic and spectroscopic criteria.
   We use high resolution echelle spectroscopic observations of the sample to
 measure accurate radial velocities (RVs).
 Distances are calculated and compared to those of the moving group from 
 the literature, we also calculate the kinematic
 Galactic components (U,V,W) of the candidate members and apply kinematic criterion of membership
 to each group. In addition we measure rotational velocities ($v\sin{i}$) to place
 further constraints on membership of kinematic members.
   We find that 49 targets have young disk kinematics and that
   36 of them possibly belong to one of our five moving groups.
  From the young disk target objects, 31 
  have rotational velocities
  in agreement with them belonging to the young disk population.
 We also find that one of our moving group candidates, 2MASS0123-3610, 
 is a low-mass double lined spectroscopic binary, with probable spectral types
 around M7.

\end{abstract}

\begin{keywords}
stars: low-mass, brown dwarfs -- stars: kinematics
\end{keywords}

\section{Introduction}
Ultracool dwarfs (UCDs) include objects with spectral type of M7 or later
(corresponding to Teff$<$2500K). Many of these objects have been identified
 in recent and ongoing large area surveys
 (i.e. 2MASS, SDSS, UKIDSS). However, the physics of their cool
 atmospheres, as well as their kinematic properties are not yet fully known.
 UCDs have complex atmospheres, the spectra of which are dominated
 by molecular bands and dust. Theory and observation
clearly suggest that the spectra of these objects are significantly
 effected by gravity and metallicity (e.g. Knapp et al. 2004, Burgasser
 al. 2006), but current model spectra cannot fit these parameters
 (Lyubchik et al. 2007). UCDs
 whose properties can be inferred without atmospheric models would give
  us rigorous test-bed for theory and vital information about how
 the physical properties affect the spectra. 

 One variety of benchmark UCD is in binary systems, since UCD
 properties may be constrained by the primary star
 (e.g. Day-Jones et al. 2008; Zhang et al. 2010; Burningham et al. 2010).
 Other way to obtain such {\it benchmark} objects is to 
 find members of known, characterised moving groups.

 A classical moving group (MG), as defined by Eggen (1984a,b),
  is a young stellar population that shares a common 
 space motion (e.g. Pinfield et al. 2006). MGs 
 form when the gas remnants of a star-forming cloud are cleared (by hot star 
 action) and many of the stars become unbound.  This population slowly disbands 
 (at a few kms$^{-1}$), but maintains its bulk space motion for up to $\sim$1 Gyr 
 before being dispersed by disk heating mechanisms (De Simone et al. 2004). MGs may 
 leave behind a bounded core of stars that become open clusters, so young open clusters 
 are often in MGs. Because of their common origin, MG members have a shared age 
 and composition, which can be well constrained by studies of high mass members
  or associated open clusters (Barrado Y Navacu\'{e}s et al. 
 1999). MGs thus represent young coeval populations with well constrained ages, and 
 because they are dispersed through space, they can be relatively nearby.  
 These nearby UCD young MG members are also optimal targets for faint companion searches, in particular,
 for high resolution imaging searches where the more manageable brightness ratio between 
 primary and secondary would favour identification (e.g. Jenkins et al. 2010).

 However, recent studies, (Famaey et al. 2007, 2008; Antoja et al. 2008; Klement et al.
 2008; Francis \& Anderson 2009; Zhao et al. 2009; L\'opez-Santiago et al. 2009),
 seem to support a dynamic or ¿resonant¿ mechanism origin of MGs or find a high percentage of
 contamination of old field stars.
 Mayor (1972) and Kalnajs (1991) suggested a different dynamic origin of MGs.
 Dehnen (1998) pointed out that the origin of most MGs could be caused by orbital resonances
 and Skuljan et al. (1999) related the origin to the Galactic spiral structure, or to some other
 global characteristic of the Galactic potential, combined with
 the initial velocities of the stars.
 But both hypothesis, that kinematic groups are cluster remnants or that
 MGs origin are dynamic, are compatible.

  Some authors have studied the contamination of young MGs by older field stars.
 L\'opez-Santiago et al. (2009) analysis of a late type kinematic member sample
 indicates that the contribution of old field stars to the sample of candidates
 of Local Association MG was $\approx$50\%, with similar results
  in the Pleiades MG by Famaey et al. (2008).
  Due to this, spectroscopic complementary studies of youth signatures
  is crucial to assess the kinematic membership. The combined information
  of kinematic and spectroscopic signatures of youth will provide a confident
  member sample.

The best documented groups are the Hyades supercluster (Eggen 1992b) associated with the Hyades 
 cluster (600 Myr), the Ursa Mayor group (Sirius supercluster) (e.g. Eggen 1984a, 1998b,
  Soderblom \& Mayor 1993a, b) associated with the UMa cluster (300 Myr), 
 IC 2391 supercluster (35-55 Myr) (Eggen 1991, 1995),
 the Castor Moving Group (200 Myr) (e.g. Barrado y Navascu\'es 1998), and the Local Association
 (20-150 Myr) or Pleiades moving group (a coherent kinematic stream
 of young stars with embedded clusters and associations such as the Pleiades,
 $\alpha$ Per, IC 2602, etc) (Eggen 1983, 1992c).

 In this paper we present the second stage of a major survey of UCDs in moving groups
 (building in the work presented in Clarke et al. 2010, hereafter Paper I). We centre here on 
the results obtained in a study of
 the candidate MG members using high resolution spectra to access their kinematic 
 signatures and resulting MG membership criteria.

In Section~2 we describe the sample selection. 
In Section~3 we give the details of our observations and our data reduction techniques.
 In Section~4 we present the new parameters obtained for our sample. In Section~5 we 
 assess our candidates properties in the context of MG membership. Finally, 
 in Section~6 we give a brief summary and the near future prospects for improved measurements.

\section{Sample Selection}
The sample presented here was selected from Paper I where we used photometric
 and astrometric criteria in order to search for UCD in moving groups.
We combined an extended version of the Liverpool-Edinburgh High
 Proper Motion survey (ELEHPM) and Southern Infrared
Proper Motion Survey (SIPS) with the
Two Micron All Sky Survey (2MASS), and applied colour cuts of
J-K$_s\geq$1.0 and R-K$_s\geq$5.0 to select objects with spectral
type predominantly later than M6.
 We selected objects from both catalogues where $\mu/\Delta\sigma$$_{\rm{\mu}}>4$
($\mu$ is proper motion and $\sigma_{\rm{\mu}}$ is the proper motion
one sigma uncertainty) to ensure proper motion accuracy.

Our final red object catalogue was made up of 817 objects.
 By using both astrometric and photometric criteria, we concluded that
  132 of these objects were possible members of one or more of the five MGs in study.
 The application of these criteria is detailed in Paper I.
 We perform high resolution follow up of a representative subset of this sample,
 all with spectral types in the range M4.5-L1 and with $\approx$45\% of the sample being M7 or later.

\section{Observations and reduction}
In this paper we analysed optical and near infrared
 high resolution echelle spectra of 68 objects. The data
 were obtained during four observing runs detailed in Table~\ref{tab:obs}.

The UVES (Dekker et al. 2000) spectra were extracted using the standard
reduction procedures in the
IRAF\footnote{IRAF is distributed by the National Optical Observatory,
which is operated by the Association of Universities for Research in
Astronomy, Inc., under contract with the National Science Foundation.}
 echelle package (bias subtraction, flat-field division,
 extraction of the spectra, telluric correction and calibration).
  We obtained the wavelength calibration by taking
 spectra of a Th-Ar lamp. The average signal-to-noise of the data, measured as 
 the square root of the signal, in these runs is $\approx$22. 

Whilst the wavelength coverage of these runs is ample, low
signal-to-noise due to poor efficiency at the echellogram ends renders
portions of the spectrum unusable. As a result, it was not possible to
measure certain important features such as the Li {\sc I}
$\lambda$6707 \AA\ absorption line.

 A similar procedure was followed for the Phoenix data. The observations in this
 case were made in a set of sub-exposures in an ABBA jitter pattern to
 facilitate background subtraction. Due to the
 lack of Th-Ar lamp, the wavelength calibration was done using sky lines
 in every spectrum with sufficient signal in the sky. For the objects 
 with low signal in the sky, we used the spectra of a giant star 
 (HR 5241).  The average signal-to-noise of the
 data in this run is $\approx$15.

 In the case of FEROS (Kaufer et al. 1999)
 data, Starlink software was used for bias subtraction,
 flat-field division and optimal spectra extraction.
 The extracted spectra were wavelength calibrated using
 the Th-Ar illuminated images obtained after each observation.
  The average signal-to-noise of the
 data in this run is $\approx$25.
 For details of our FEROS reduction see  Paper I.

\section{Measurements of parameters}

\subsection{Spectral Types and distances}

 We performed a simple analysis to obtain the objects' spectral types.
  We measured the PC3 index as defined in Mart\'{i}n et al. (1999)
 and applied the index-spectral type relation to derive a spectral type.
 We measured also TiO5 and VO-a indexes as defined in Cruz \& Reid (2002).
  We derived spectral type from TiO5 following the double valued relation given
  in Cruz \& Reid (2002), using the relation for Ti05$<$0.75, except for object
  with spectral type later than M7 were we used the relation given for Ti05$>$0.30. 
 We also derived a spectral type from VO-a index following the relation gives
  in Cruz \& Reid (2002) for the correspondent spectral type interval.
 We took as definitive spectral type the average of these three spectral types
 rounded to the closest 0.5 class.
 When one of the spectral areas covered by these indexes were no usable
 we make the average between the other two.
 The spectral type estimated error is $\approx$1.0 spectral class 
 derived from these relations.         

 Due to the wavelength range of the Phoenix data, we can not measure PC3, so
 we measure the spectral types instead using the I-J colour index-spectral type relation
 from Leggett (1992).

 We then calculated a more accurate spectroscopic 
 distance using a PC3-absolute J-magnitude relation from Crifo et al. (2005).
 The error in distance following Crifo is 12\%.
 For Phoenix data, distances were calculated from the absolute 
 J magnitude-spectral type relation from Dahn et al. (2002).
 The spectral classification and distances are given in Table~\ref{tab:par},
 where previous known data from the literature is also given. The values in 
  both sets are in agreement except for two objects, 2MASS J0334-2130 and
 SIPS 1632-0631, that differ in more than one spectral type and 
 SIPS2039-1126 that differ one spectral type. 
 The value from the literature for
 2MASS J0334-2130 and SIPS 2039-1126 classification comes 
 from Cruz et al. (2003), where all spectral types 
 for M dwarfs were determined via visual comparison with standard star
 spectra taken during the course of the program (the
 objects were typed by being normalised and plotted between
 spectra from a grid of eight standard M1-M9 dwarfs
 from Kirkpatrick, Henry \& McCarthy 1991). In this case we think 
 our index average measurements will give more accurate spectral types.
 For SISPS 1632-0631, we took the spectral type from Gizis 
 (2002), where the final classification  comes from the measurements 
 of several spectral indexes (including TiO5, PC3,
 CrH, TiO-b and visual inspection). The PC3 and TiO5
 in Gizis (2002) (1.99, 0.25) are very similar to the ones calculated by us (1.95, 0.26),
 but our value of VO-a (2.35) is the double of Gizis's (1.14).
 Although Gizis (2002) takes an average of six indicators while we only take three,
 we rely on our classification.
 As we mentioned above, the spectral type-index relation of TiO5 and VO-a indexes in the Cruz \& Reid
 (2002) is double value with a turning around M7 and M9 for TiO5 and VO-a respectively,
 that is that there is a different relation for objects with spectral type under or over around
 M7 for TiO5 and M9 for VO-a. SIPS 1632-0631 is just in the M7-M9 limit. We find that
 the best agreement in the spectra type with PC3 is when we used the relation for $>$M7 in TiO5 and 
 for $<$M9 in VO-a, giving same spectral type, M8.5, for the three indexes.

 Due to the spectroscopic binarity, as we will see in Sect. 5.3 or possible binarity 
 without confirmation of some object in the 
 sample we expect that the error in the indexes spectral types and 
 distances are larger than the ones shown in Table~\ref{tab:par}.

 To obtain spectral types for our FEROS objects we initially used
  the PC3 index from Mart\'{i}n et al. (1999), however this 
 coincided often with a unusable region of spectra. 
  We thus used the VO index from Kirkpatrick,
  Henry \& Simons et al (1995), CaH from Kirkpatrick, Henry \& McCarthy (1991),
  VO-a and TiO5 from Cruz \& Reid (2002) and PC3, TiO1+TiO2 and 
  VO1+VO2 from Mart\'{i}n et al. (1999) to create a mean spectral type  
  (see  Paper I for details).

\subsection{Radial velocities}
The heliocentric radial velocities
have been determined by using the cross-correlation technique.
The spectra of the targets were cross-correlated order by order, by
using the routine {\sc fxcor} in IRAF, against spectra of radial velocity
standards with similar spectral type.

The radial velocity was derived for each order
from the position of peak of the cross-correlation function peak (CCF),
and the uncertainties were calculated by {\sc fxcor} based on the
fitted peak height and the antisymmetric noise as described by
Tonry \&  Davis (1979).

In the case of the FEROS data, we cross-correlated each
 individual order of the reduced spectra using an IDL
routine that determines the cross-correlation of
two arrays and finds the maximum cross-correlation
function by fitting a Gaussian to the central peak.
  Our final radial velocities
 and associated uncertainties came from the mean
and standard deviation of these measurements.
 A heliocentric correction was made to the measured
 shift between the object and the reference star so
that the we could calculate radial velocities with
respect to the Sun.

In Table~\ref{tab:ref} we give the information about the reference stars used in
 every run. The object LP 944-20 exhibits $\sim$3.5 km s$^{-1}$ radial
 velocity variability, as is shown in  Mart\'in et al. (2006), so when we had another
 radial velocity template in the run we checked the velocity of LP 944-20 with respect to
this other template and found good agreement within the uncertainties.

In Table~\ref{tab:vrs} we list, for each spectrum, the
heliocentric radial velocities ($V_{\rm r}$)
and their associated uncertainties ($\sigma_{V_{\rm r}}$)
obtained as weighted means of the individual values deduced for each order used.
Those orders which contain chromospheric features and prominent
telluric lines have been excluded.
In the case of Phoenix data, the signal-to-noise of the spectra is low
 and so we decided to add an uncertainty 
 deduced from the variation found in the standard stars calibrations
 (LP 944-20 and GL 406 in this case).

For the binary case (2MASS0123-3610), we have to take into account an additional
 uncertainty due to the systematic distortion
 of the peak from the presence of a companion.

\subsection{Space motions}

 To constrain space motions, we obtained a 
 number of distances depending of potential MG membership
 for each object with radial velocity measure.

 As detailed in  Paper I, to allow for younger objects
 appearing intrinsically brighter, we created isochrones
 corresponding to the ages of our MGs and that of a field
  dwarf using atmospheric models from Baraffe et al. (1998).
 According to each object's candidature group membership from photometric
 and astrometric criteria, the difference in absolute J-magnitude of each object
 is used in Dahn et al. (2002) distance-spectral type relation to give 
 a new corrected distance assuming group membership.
 This way we obtained a number of distances for each of the
 objects.

 Using these distances or any parallaxes
 available from the literature, the proper motion data, and the radial
 velocities calculated in Sect. 4.2, we computed the Galactic space-velocity
 components ($U$, $V$, $W$) of the sample by using the
 transformation matrices of Johnson \& Soderblom (1987).
  Positive U is towards the Galactic
 centre, positive V is in the direction of Galactic rotation, and
 positive W is in the direction of the Galactic north pole.

 The resulting values of ($U$, $V$, $W$) according to potential group membership
  and associated errors are given in Table~\ref{tab:vrs}.
 In the table, the objects that appear to be kinematic members,
 that is, that are in agreement with the membership,
 are highlighted in bold (see Section 5).

\subsection{Rotational velocities}

To determine rotational velocities we made use of
the cross-correlation technique in our high resolution echelle
spectra by using the routine {\sc fxcor} in IRAF.

When a stellar spectrum with rotationally broadened lines is
cross-correlated against a narrow-lined spectrum, the width of the
cross-correlation function (CCF) is sensitive to the amount of
rotational broadening of the first spectrum.
Thus, by measuring this width using Gaussian profiles, one can obtain a measurement of
the rotational velocity of the star.
The target spectra were cross-correlated against the
spectrum of the template stars. We give the data of the templates used in
 each run in Table~\ref{tab:ref}.
The templates should be of equal or similar spectral types as the targets,
 but an M-type template can be used for all M type dwarfs (see e.g.
  Bailer-Jones 2004; Zapatero-Osorio 2006), therefore as our targets and references
 are all in M4-M9 range, we applied our templates to all stars.

 The calibration of the width (FWHM) of the templates
 to yield an estimation of $v\sin{i}$ is determined by
 cross-correlating artificially broadened spectra of the template star
 with the original template star spectrum. We rotate them up to 100 km s$^{-1}$
 in 1 km s$^{-1}$ steps. The broadened spectra were created for $v\sin{i}$ is spanning the
 expected range of values by convolution with a theoretical rotational
 profile (Gray 1992) using the program {\sc starmod}
 (developed at Penn State University; Barden 1985) and
 the {\it vsini} routine of the {\sc SPECTRUM} synthesis code
 (Richard O. Gray, 1992-2008).
 The resultant relationship between $v\sin{i}$ and FWHM of
 the CCF was fitted with a fourth-order polynomial. 
 This method should not be used at rotational velocities greater 
 than around 50 km s$^{-1}$ since the CCF profiles radically differ from Gaussianity.
  Therefore, in our rotational range of interest, 5$>$$v\sin{i}$$>$50, the fit is good.
 We used several spectral ranges for these measurements.
 As mentioned in Reiners \& Basri (2008), absorption band of FeH around 1$\mu$m
 is a rich area in ultracool stars, free of telluric absorption
 and is not pressure-broadened, but the spectra that we had covering that area
 (UVES data) present an artifact that do not allow us to make any measurements with
 the reliability needed. So, we use that band only to check the values obtained
 in other areas.

 The rotational velocities given in the 8th column
 of Table~\ref{tab:vrs} are the average of the values
 measured when more than one template star was used.

 The uncertainties on the $v\sin{i}$ values obtained by this method
 have been calculated using the parameter R defined by Tonry \&  Davis (1979)
 as the ratio of the CCF height to the rms antisymmetric component.
 This parameter is computed by the task {\sc fxcor}
 and provides a measure of the signal-to-noise ratio of the CCF.
 Tonry \&  Davis (1979) showed that uncertainties in the FWHM
 of the CCF are proportional to 1/(1 + R) and
 Hartmann et al. (1986) and Rhode et al. (2001) found that the
 quantity  $v\sin{i}$[1/(1 + R)] provides a good estimate
 for the 90\% confidence level of any $v\sin{i}$ measurement using this methodology. This error, $v\sin{i}$[1/(1 + R)], should be a
 reasonable estimate of the uncertainties on our $v\sin{i}$ measurements.
 Nevertheless, we estimate the uncertainties to be at least 3 km s$^{-1}$
 for all values, which is the level of variation presented by the 
 reference stars. The
 reason is that the main source of uncertainty comes from the low signal-to-noise 
 of the spectra and so is not reflected properly in the statistical
 results of the calculations.

Taking into account the resolution of our spectra, the rotation of our
 templates (see Table~\ref{tab:ref}) and the signal-to-noise, we consider that
 the minimum detectability value
 of $v\sin{i}$ would be in the range $\approx$6-10 km s$^{-1}$. Thus, we have assumed
 a general limit of 10 km s$^{-1}$. All data below this limit
 would be marked as $v\sin{i}$ $\le$ 10 km $s^{-1}$ in Table~\ref{tab:vrs}.

\section{Analysis}

\subsection{Space motions and kinematics}

We apply a simple kinematic criterion
 to distinguish old and young disk objects in our sample of 68 objects
 with high resolution spectra.

\begin{figure*}
   \centering
   \includegraphics[width=8.5cm,clip]{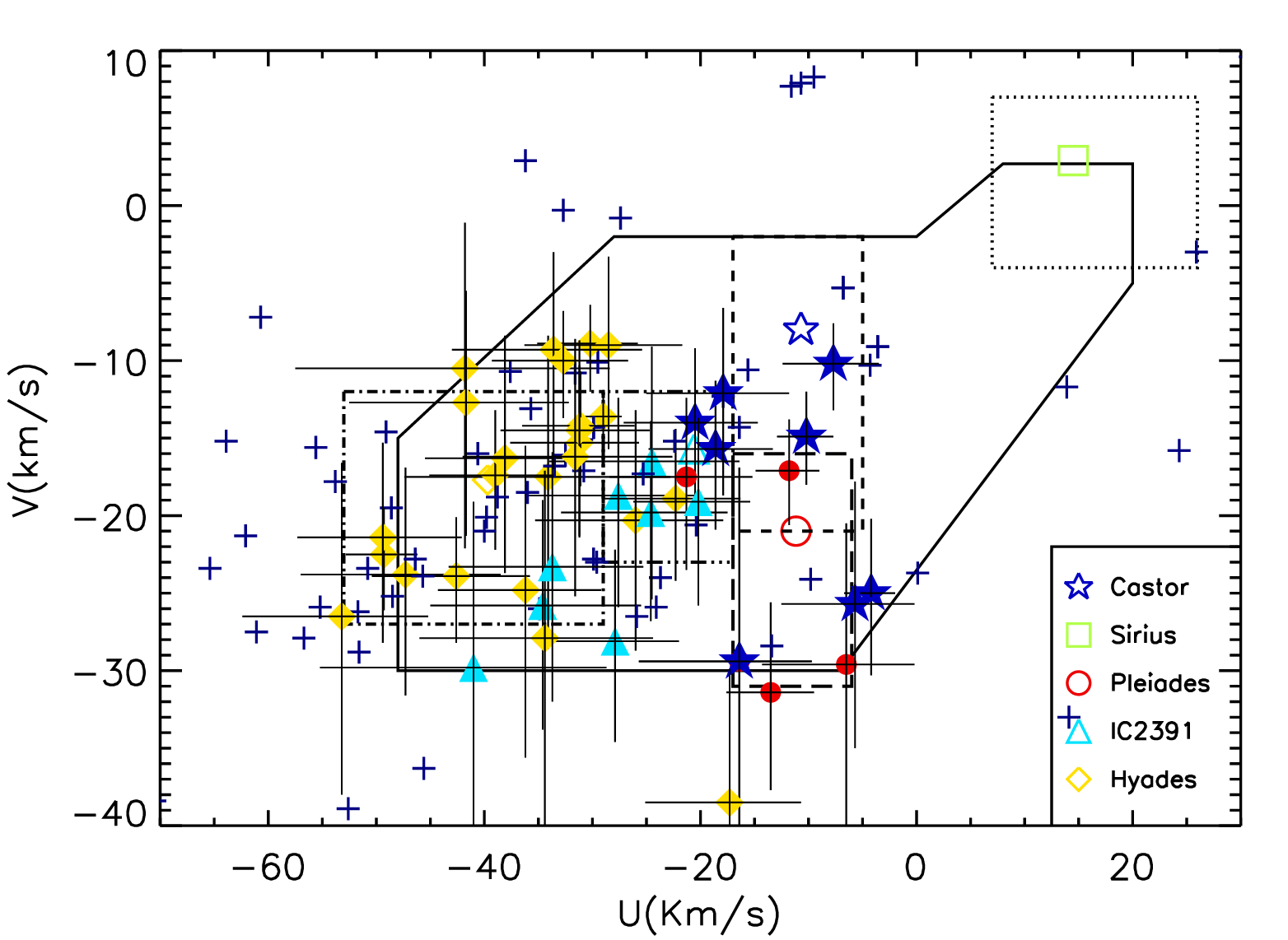}
   \includegraphics[width=8.5cm,clip]{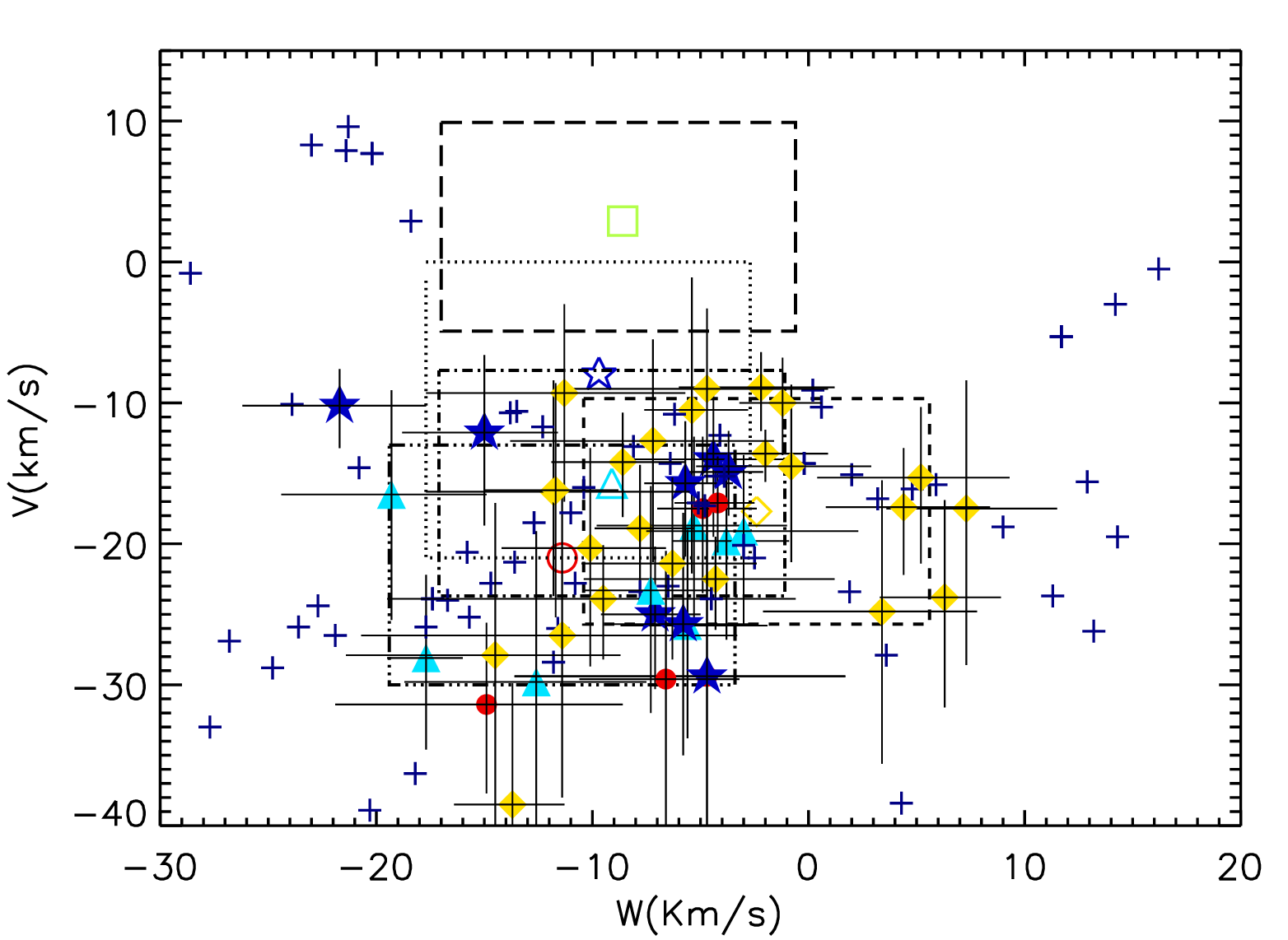}
      \caption{
 We plot here the UV and VW space motion diagrams for our sample.
 The young disk boundaries defined by Eggen (1984b,1989) are plot as continuous
 line. Objects inside these boundaries might be young (YD).
 The boxes indicate the expected kinematic range for the five MGs.
 The centre of the 5 main MG are marked (with large open symbols) and
 our kinematic members are shown as corresponding filled symbols.
 The objects that are not candidates to be in MGs are plot
 in crosses. 
 A kinematic member was defined lying inside the relevant box
 within the uncertainties.
              }
         \label{fig:fig1}
   \end{figure*}

In Figure~\ref{fig:fig1}, we plot the UV and WV planes for the whole sample
  including the boundaries (continuous line)
 that determine the young disk population as defined by Eggen (1984a,b, 1989).
   We then classify objects as  young disk (YD) if they lie within the
   young disk region in the UV plane (or if they could, given U and V
   1$\sigma$ errors). Objects whose location in the UV plane is
   inconsistent with YD membership are all consistent with an old disk (OD)
 classification.

 If we follow Legget (1992) criterion, the sample should be divided into
 young disk (YD), old disk (OD), young-old disk (YO), old-disk halo
 (OH) and halo (H) objects: all stars with V $<$ -100, or with an
 eccentricity in the UV plane $>>$0.5, were defined as H;
 objects with an eccentricity in the UV plane  of $\approx$0.5 were
 defined as OH; stars within the young disk region in the UV plane,
 approximately defined by -20 $<$ U $<$ 50, -30 $<$ V $<$ 0, and with
 -25 $<$ W $<$ 10, were defined as YD; objects with an eccentricity in
 the UV plane less than 0.5, that lie outside the young disk region, or
 that lie within the young disk region but have $|W|$ $>$ 50, were defined
 as OD. Finally, objects that lie around the edge of the young disk region
 or lie  within the area with $|W|$ $<$ 50 but greater than that
 required for classification as YD, were defined as YO.
 However, in our sample all the objects
 would be classified by young disk or old-young disk membership,
 (see Fig.~\ref{fig:fig1} and W value in Table~\ref{tab:vrs}), and
 will refer to them as YD and OD for simplicity.

 In addition we identify possible members of the five MGs
 by their relative position in the UV and VW diagrams (Fig.~\ref{fig:fig1}) 
 with respect to the
 boxes (dashed lines) that mark the velocity ranges of each MG.
  The centre of the five main MGs are marked (with large open symbols) and
 our kinematic members are shown as corresponding filled symbols.
 The boxes are defined by studies with samples of higher mass members
  from the literature (Barrado Y Navacu\'{e}s 1998; Montes et al. 2001).
 See  Paper I for a detailed description.
 As mentioned above, a kinematic member was defined as either
 lying inside the relevant box
 in each diagram, or having uncertainties that overlapped appropriately.
 Objects inside YD boundaries should be young although it
 is not clear that they belong to an established MG.

According to their kinematic behaviour, we have 49 objects
 that belong to the young disk area, and 36 of them are
 candidates members of one of the five MGs. Some objects are
 candidates for more than one MG (see Table~\ref{tab:criteria}).
 We find seven possible candidates to Castor, 25 to Hyades,
 seven to IC 2391, five to Pleiades and none to Sirius.
 Eleven objects belong to young disk region without clear
 membership to any of these five MGs.

These new candidates represent a significant addition to the low-mass
membership of several MGs that for the most part have few low-mass
members confirmed to date, and provide a corresponding improvement in
any determination of the low-mass initial mass function (IMF). Our
seven candidate members of the Castor MG represent an increase of 10\%
in the total known membership (see e.g. Caballero 2010), and an increase of
50\% in the very low-mass regime.
The Hyades group presents a deficiency of very low-mass objects (15 of
$\approx$500 stars). Bouvier et al (2008) suggest that the Hyades originally
had 150-200 brown dwarfs (BDs) compared to its current 10-15, based on the
assumption that the Hyades is a dynamically evolved version of the
Pleiades and considering the two clusters present mass functions over
the 0.005-3M$_{\odot}$ range. Hogan et al (2008) reported 12 new L dwarf Hyades
members (estimating $\approx$20\% contamination). Here we present 25 low-mass
candidates which, if confirmed, will further aid our understanding of
the evaporation of young clusters.
 IC 2391  as well as other young clusters present also a brown dwarf deficit,
 that might be explained by an onset of larger-size dust grain formation in the upper
 atmosphere of objects with spectral types M7-M8 or later (Dobbie et al. 2002;
 Jameson et al. 2003; Barrado Y Navacu\'{e}s et al. 2004; Spezzi et al. 2009). 
 From 180 members around a dozen are
 substellar (Spezzi et al. 2009), a low number from which to draw a statistically
 significant conclusion. Although our candidates are M4.5-M7 spectral type, they
 increase the low-mass population and favour future statistics.
 Pleiades is a very well studied and populated cluster and many works
 have undertook the low-mass and BD population search and IMF study (e.g.
 Londieu et al. 2007), our five candidates are in M6-M8.5 spectral type
 and will contribute to populate this spectral type range.

 We have to keep in mind that contamination of the young disk space velocity area
 by old field population occurs. It has been studied in several kinematic
  moving groups as we mentioned in the introduction, concluding that a 
 high percentage of contamination happens (e.g. L\'opez-Santiago et al. 2009 and reference therein).
 So the next logical step is to further asses the kinematic candidates 
 found with additional information, such as the study of several youth
 indicators, to confirm their membership.

\subsection{Rotational velocity-spectral type relation}

 The rotational evolution of low-mass object is complicated.
 Several studies have demonstrated that after becoming a fully convective
 object (spectral types later than  $\approx$M3), when gravitational
 contraction is finished, rotational braking is present at least in 
 all M type objects. The spin-down times are longer for Late M dwarfs than
 for earlier M types (e.g. Zapatero-Osorio et al. 2006; 
 Reiners \& Basri 2008, Jenkins et al. 2009). 
 Jenkins et al. (2009) show a statistical 
 relationship between mid/late type M dwarf rotation and age and hence 
 taking all these works together we can use rotational velocity as a way to 
 differentiate between young and older M type UCD populations.
 As demonstrated
 in Figs. 9 and 10 of Reiners \& Basri (2008), such populations occupy
 distinct regions of the $v\sin{i}$-spectral type diagram depending on its age, 
 with the older objects concentrated in an "envelope of minimum rotation velocity".
 Even for the latest M dwarfs this envelope corresponds to a fairly
 low rotation rate ($v\sin{i}$ $<$ 20 km s$^{-1}$). The model overlays 
 that Reiners \& Basri (2008) derive (based on a breaking law fit to the overall
 observed population) that the great majority of mid-late field
 M dwarfs (e.g. with ages $\sim$2-10 Gyr) have $v\sin{i}$ $<$ 30 km s$^{-1}$, and
 that younger objects of this type should be much more rapid rotators
 ($v\sin{i}$ = 30-100 km s$^{-1}$). Although the random inclination of rotation
 axis can reduce measured $v\sin{i}$, it seems clear that rotational
 velocity constraints can provide a very helpful complement to
 our kinematic membership.

\begin{figure*}
   \centering
   \includegraphics[width=8.5cm]{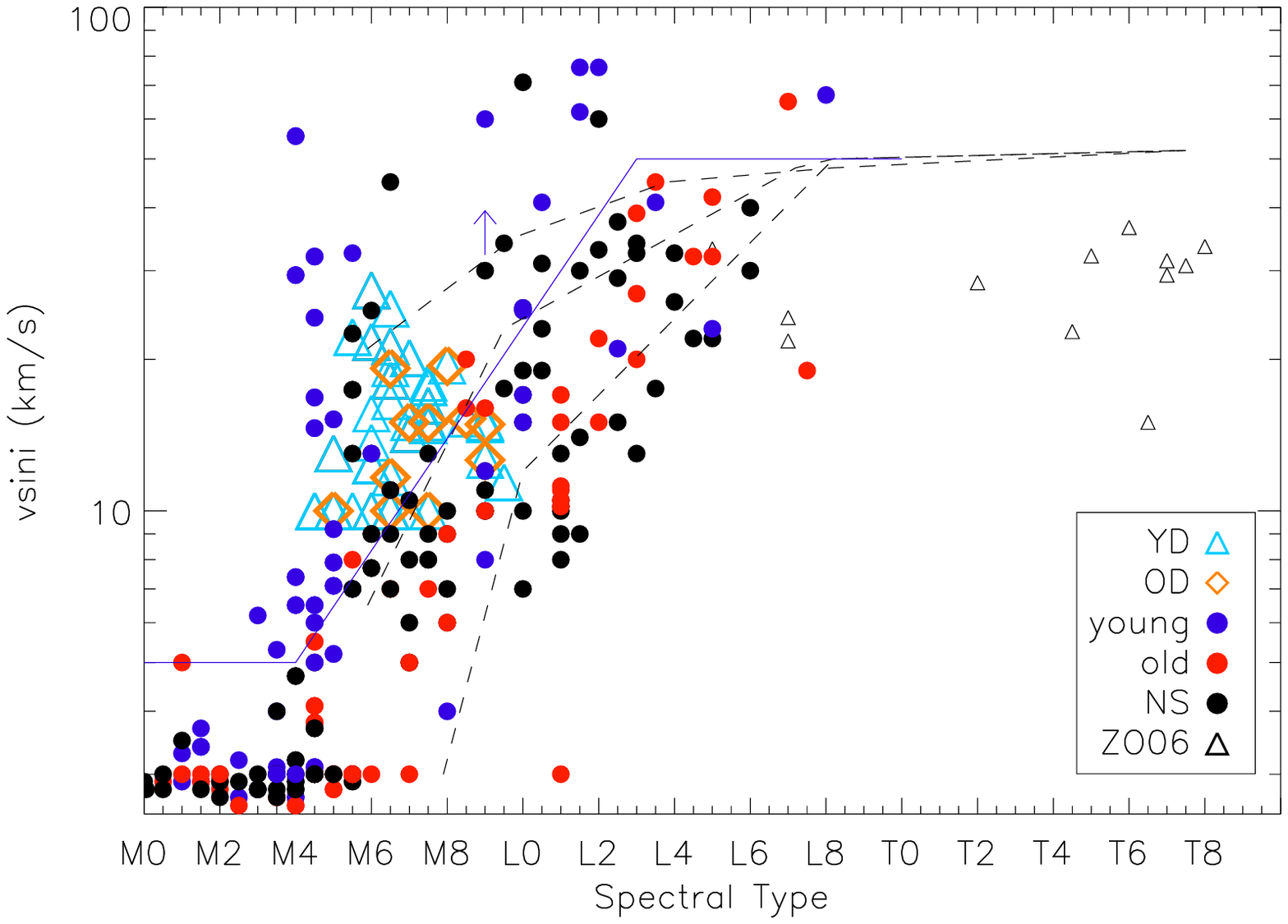}
   \includegraphics[width=8.5cm]{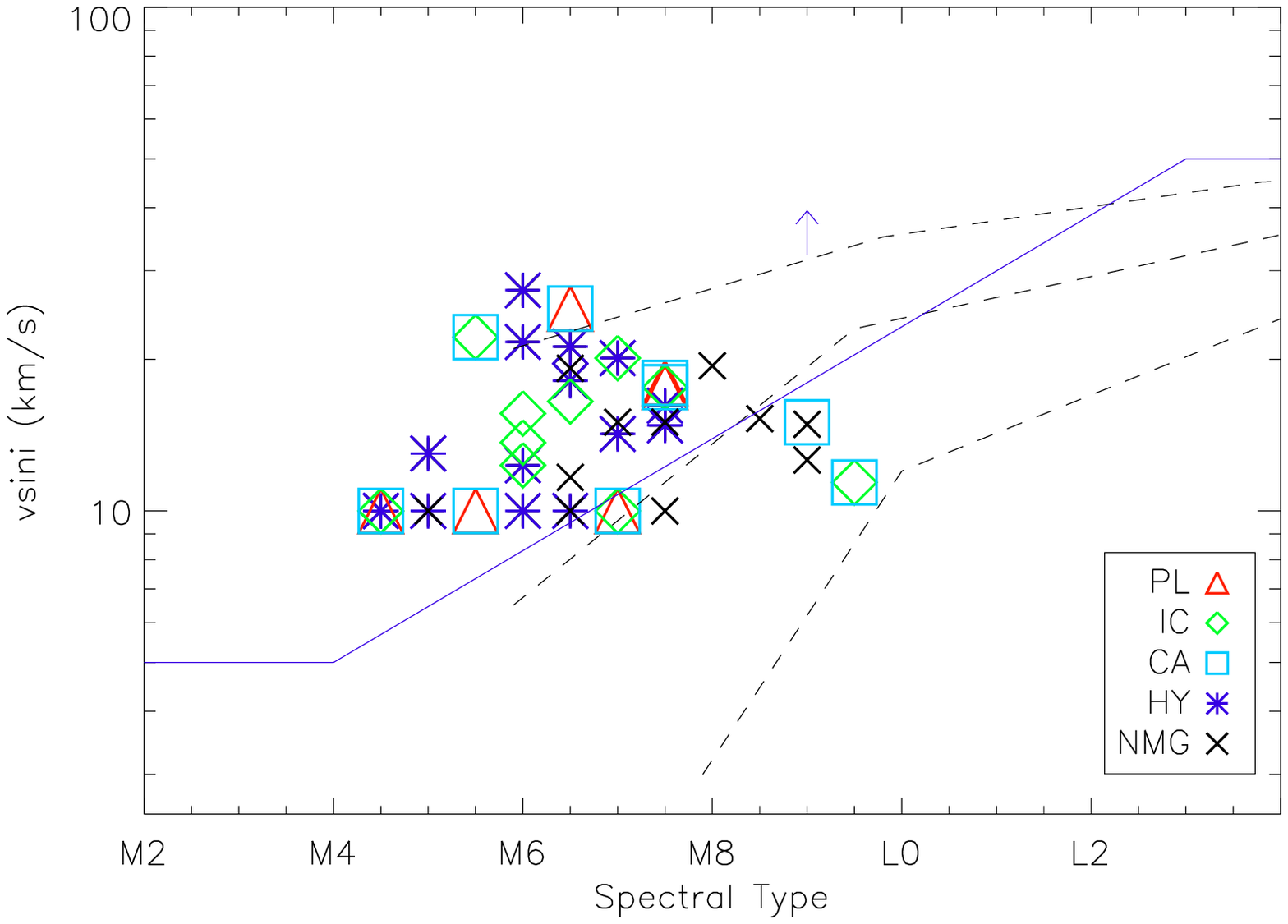}
      \caption{
Spectral type versus $v\sin{i}$. Left: we plot our sample with Reiners \& Basri (2008) figure 10
 (circles) and open triangles from Zapatero-Osorio et al. (2006).
Blue circles are supposed to be young, red circles old and black circles have
 unknown age. Our data sample is plotted as different symbols depending if 
  they have been classified as YD or OD. As in Reiners \& Basri (2008), we mark in
  dashed lines ages of 2, 5 and 10 Gyr (from upper left to lower right).
  We plot in blue continuous line the "apparent" separation between the consider "young"
 and "old" object and so use it as a criterion to determine the possible
 membership of our targets to the MG they are candidates to.
Right: our data sample in the same plot but with different symbols 
 depending of the MG they are a candidate for. The rest of literature data is not plotted here
 for clarity.
}
         \label{fig:fig2}
   \end{figure*}

One Reiner \& Basri (2008) result is that the rotational
 evolution according to a wind-braking law that scales with temperature can 
reproduce better the observational data for L dwarfs. While for M dwarfs
 both rotational evolution according to a wind-braking law that scales 
 with temperature or with mass are very similar. 
 We choose the first one here although this is a moot point since
 both models give similar results over our spectral type range. 
 In Fig.~\ref{fig:fig2} we plot 
 spectral type versus $v\sin{i}$. On the left, we plot our sample with 
 Reiners \& Basri (2008) Figure 10.
 Circles are from them and open triangles from Zapatero-Osorio et al. (2006).
 Following Reiners \& Basri (2008), filled blue circles are probably  
  young, filled red circles old and open circles have no age information.
 Our sample data is plotted as different symbols depending if they
  have been classified as YD or OD. As in Reiners \& Basri (2008), we mark in
  dashed lines ages of 2, 5 and 10 Gyr (from upper left to lower right).
 To obtain a criterion to separate objects of differing age, we plot in
  blue a continuous line: the "apparent" separation between a "young"
 and "old" object and so use it as an
 indicator of MG membership for our targets.
 We include LP 944-20 in the plot
 as it shows a strong candidature to
 be a member of Castor MG (see Ribas 2003b for details). It is
 plotted as an arrow with our measured value of 
 $v\sin{i}$, in agreement with previous works (see Table~\ref{tab:ref})
 and with being young.
 On the right side we show a zoom version where we
 plot the targets in different symbols, which indicate for which MG
 they are candidate members.
 In this case we do not plot literature data for clarity.
 Note that the objects marked as $<$10 km $s^{-1}$ in Table~\ref{tab:vrs} 
 are plotted with this value in Fig.~\ref{fig:fig2}, so this should
 mark an upper limit of rotational rate. 

We measure the rotational velocity of 54 of the 68 objects in the sample, and of 
 42 of the 49 YD objects.
 From these 42, we find that 31 present a projected rotational 
 velocity in agreement with our criterion of youth and 12 that 
 can not be dismissed as they have velocities in the limit between
 both populations or we have no rotational velocity information. 

 Despite being outside the young disk region, 10 objects
 have youthful rotational velocities. Uncertainties in the (U, V, W)
 measurements would be compounded by possible binarity among older objects 
 which can allow them to keep high rotation rates longer that in case of singles.

 Table~\ref{tab:criteria} sums up these results. The first and second
 columns are name and kinematic classification. Third column classifies
 the objects
 as young (Y) if they lie in the upper part of the limiting line,
 old (O) if they lie under the line and unknown age (Y-O) if they
 are in the limit of both areas or the rotational velocity superior limit
 is not clear enough to classify them in one of the groups. Column four
 highlights to which MG the candidates are kinematic members.
 In this case, HY for Hyades, SI for Sirius, CA for Castor, PL for Pleiades and
 IC for IC 2391, and OYD for other young disk members without a membership
 of one of the five MG in consideration. The confirmed 31 YD members  
 objects are highlighted in bold in Table~\ref{tab:criteria}, and the 12 more
   including as possible members are highlighted in bold with a question mark.
 Only two objects that presented candidature to any of the MGs have
 been dismissed due to the rotational velocity and have a (N) in
 Table~\ref{tab:criteria}.  

 As we have used the projected rotational velocity for our study, the real rotational
 velocity can only be higher and so the objects over our youth limit will
 stay there, but the dispersion introduced by the inclination and the
 fact that the rotation rate can be influenced by other external factors, like binarity,
 make us use this method only as a supporting criterion.

Five objects of our list, DENIS0041-5621, 2MASSJ0429-3123, SIPS0440-0530, 2MASSJ1507-2000 and
 DENIS2200-3038, overlap with Reiners \& Basri (2009) and Seifahrt et al. (2010).  The kinematic
 shows they are YD except for 2MASSJ1507-2000, while two of them show MG candidature, SIPS0440-0530 and DENIS2200-3038.
  DENIS0041-5621 shows evidence of accretion and presence of Li~{\sc i} 6708 \AA, which implies that it is
 very young (Reiners \& Basri 2009).  We had classified this object as other young disk, as it is kinematically
  young and the rotational velocity is in agreement.  2MASSJ0429-3123 shows no Li~{\sc i} in Reiners \& Basri (2009).
 We have this object classified as YD due to kinematics but the rotational velocity indicate it could be old,  
 classified as an M7.5 with no lithium, it is probably an old object.  SIPS0440-0530 is classified by us as 
  a young object member of Hyades MG.  Reiners \& Basri (2009) found no Li~{\sc i} but this could be compatible
    with being a HY member.  H$\alpha$ and other indicators of activity will give us more information as it 
 is M7.0 spectral type.  2MASSJ1507-2000 shows quite different radial velocity measurement in Reiners \& Basri (2009)
 (-2.5 km s$^{-1}$ instead of our -22.18 km s$^{-1}$), we can not explain this difference, the additional error maybe 
 produced by low signal-to-noise spectra is not enough for producing this discrepancy. This object is kinematically
  classified as OD and also shows no Li~{\sc i} in Reiners \& Basri (2009).  DENIS2200-3038 is kinematically
  classified as Hyades member, which is  consistent with the findings of Seifahrt et al. (2010), but with 
 M9 spectral type the rotation criterion gives  an old object. Further spectroscopic criteria is 
 needed to confirm these results.

\subsection{SB2 type binarity: 2MASS0123-3610}

Multiplicity of stars can yield dynamical mass constraints, as well 
 as being an important constraint of stellar formation
 and evolution. The properties of multiple stellar systems have long
 provided important empirical constraints for star formation theories.
 Binary properties of the low-mass stars of FGK and early M spectral
 types have been extensively studied (e.g., Duquennoy \& Mayor 1991 and 
 Fischer \& Marcy 1992). But, despite the interest in the multiplicity
 of very low mass objects (stars at the bottom of the main sequence and 
 brown dwarfs) in recent years (e.g. Mart\'in et al. 2003;
 Close et al. 2003, 2007; Gizis et al. 2003; Pinfield et al. 2003;
 Basri \& Reiners 2006; Reid et al. 2006; Ahmic et al. 2007; etc),
 and the subsequent progress in finding and characterising them,
  very few direct measurements of their physical properties have been made,
 in particular dynamical masses (e.g. Zapatero Osorio et al. 2004,
  Dupuy et al. 2009a and Dupuy et al. 2009b), and 
 our understanding of their formation processes is not clear.

  Double (or multiple) lined spectroscopic binaries (SBs) allow precise
  determination of dynamical properties of the components, including
  the mass ratio and, if the inclination can be determined, the
  individual component masses.


 During the reduction process of our sample we find some
 objects that show possible spectroscopic binarity, since 
 they show some contribution of the secondary in the spectra, 
 but we can confirm only one of them. We present here an object 
 that can be clearly defined as double line spectroscopic binary or SB2.

\begin{figure*}
   \centering
   \includegraphics[width=8.5cm,viewport=0 0 538 282,clip]{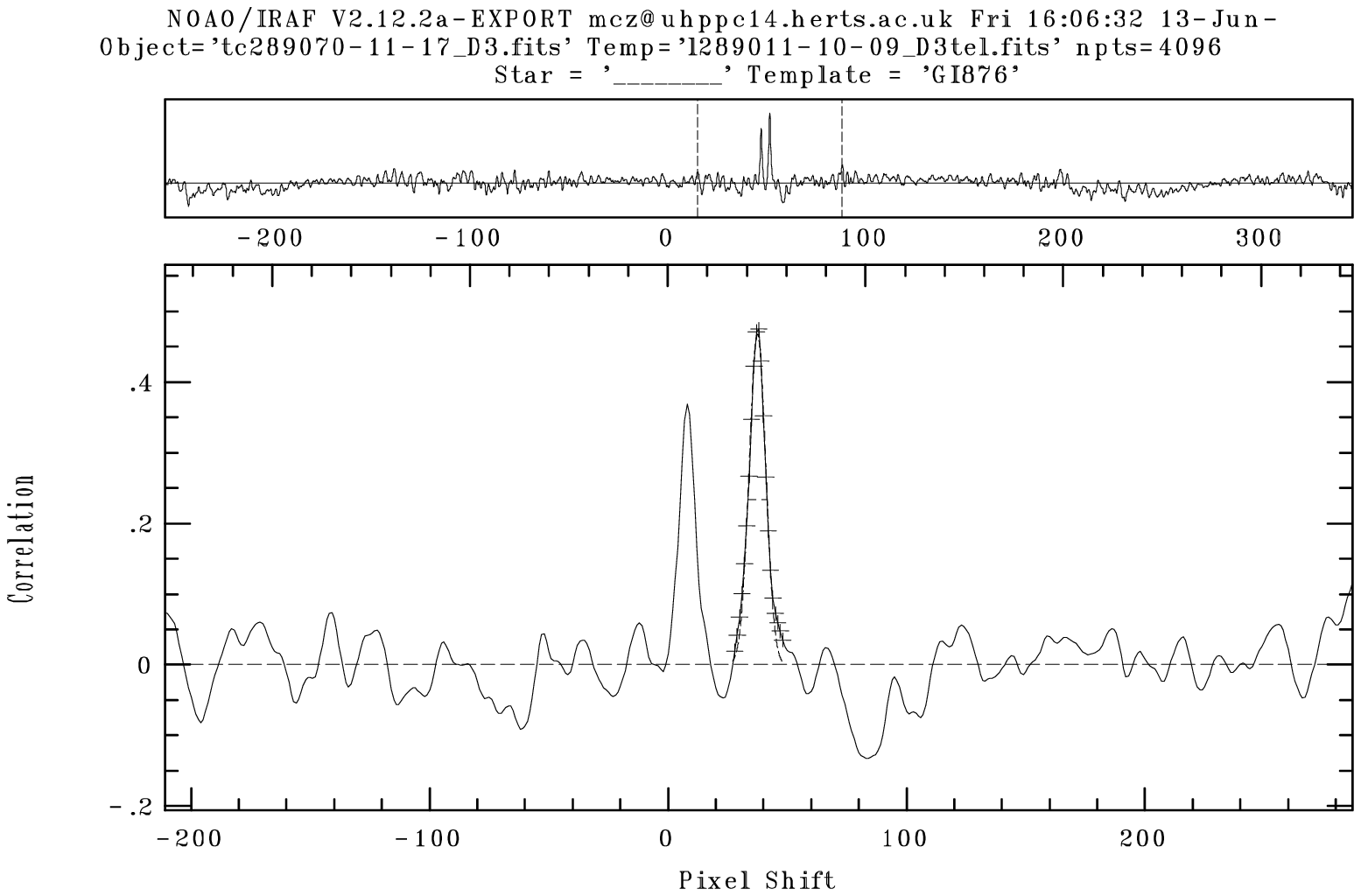}
   \includegraphics[width=8.5cm,viewport=0 0 538 288,clip]{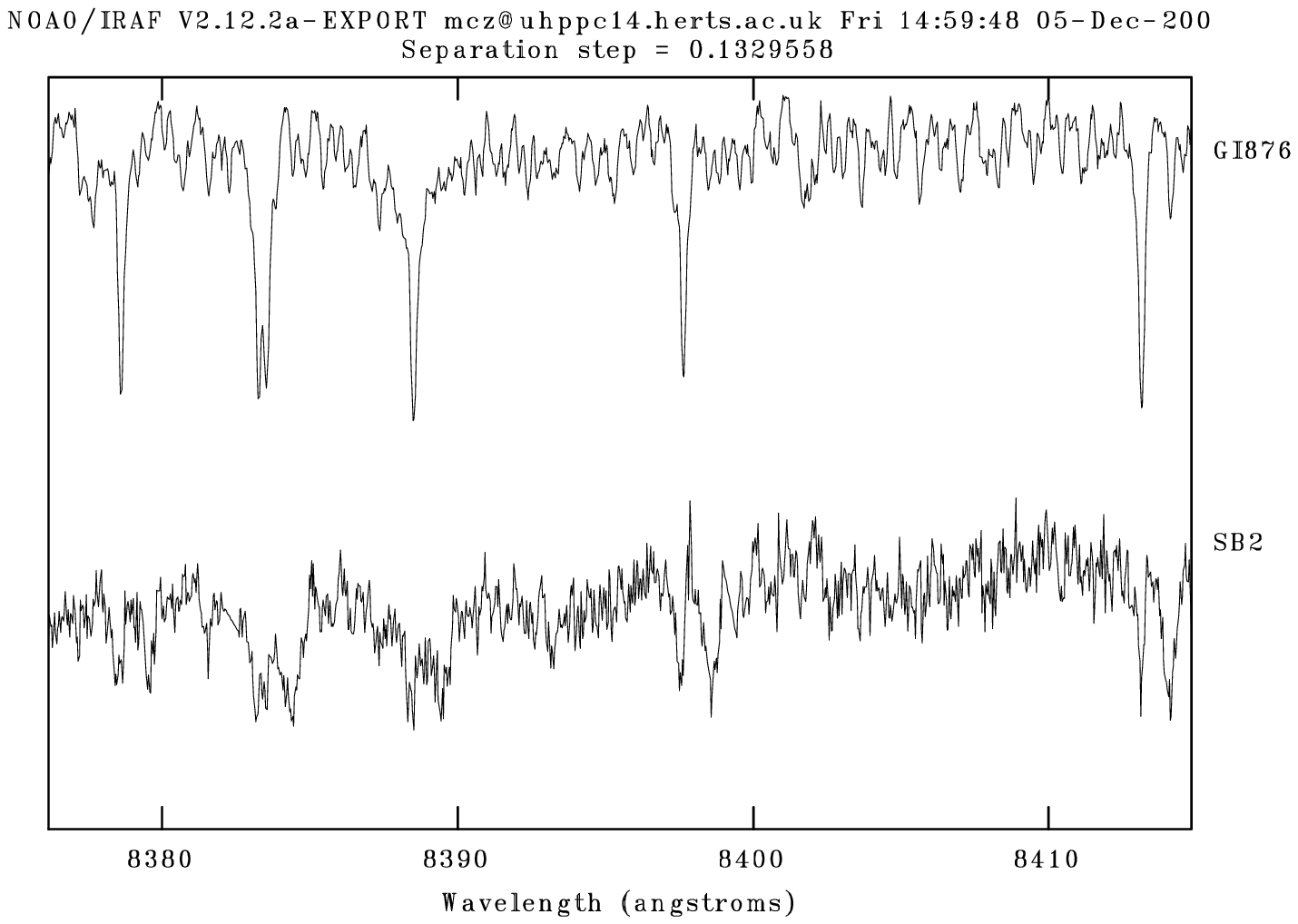}
      \caption{2MASS0123-3610: On the left side, an example of
 the  {\sc fxcor} cross-correlation where the the peak
 of each component can clearly be seen. On the right side,
 we plot a piece of spectra of our target (below) and the
 reference single star GI 876 (above). The absorption lines from both
 components are also well distinguished.
}
         \label{fig:fig3}
   \end{figure*}

On the left-hand side of Fig.~\ref{fig:fig3} we plot a example of
 the {\sc fxcor} cross-correlation were the peak
 of each component can clearly be seen and
 so fitted separately. On right-hand side of the Fig.~\ref{fig:fig3},
 we plot a piece of spectra of our target (below) and the
 reference single star GI 876 (above). The absorption lines from both
 components are also well distinguished.

Due to the SB2 nature we could measure the radial and rotational 
 velocity of both components (see Table~\ref{tab:vrs} and Sect. 4).
 We derived Galactic space-velocity components ($U$, $V$, $W$)
 in the same way we specify in Sect. 4. 

 Since we have the double lines in the spectra, 
 when we applied the indexes criteria for obtaining the spectral classification,
 we have the lines of both components together that we could not disentangle. 
 This produce an uncertainty in the spectral classification an 
 so in the distances derived.

\begin{figure*}
   \centering
   \includegraphics[width=8.5cm]{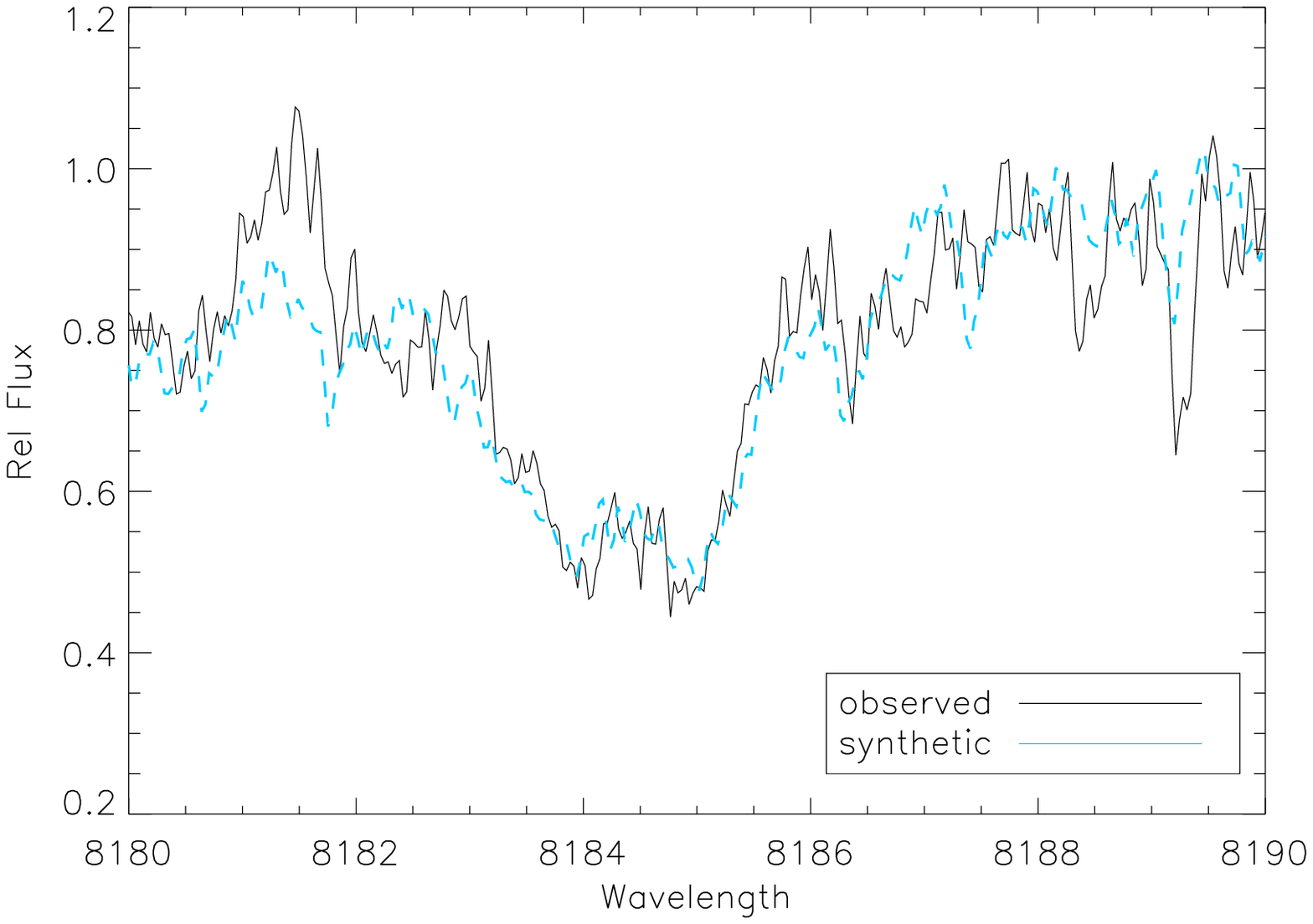}
   \includegraphics[width=8.5cm]{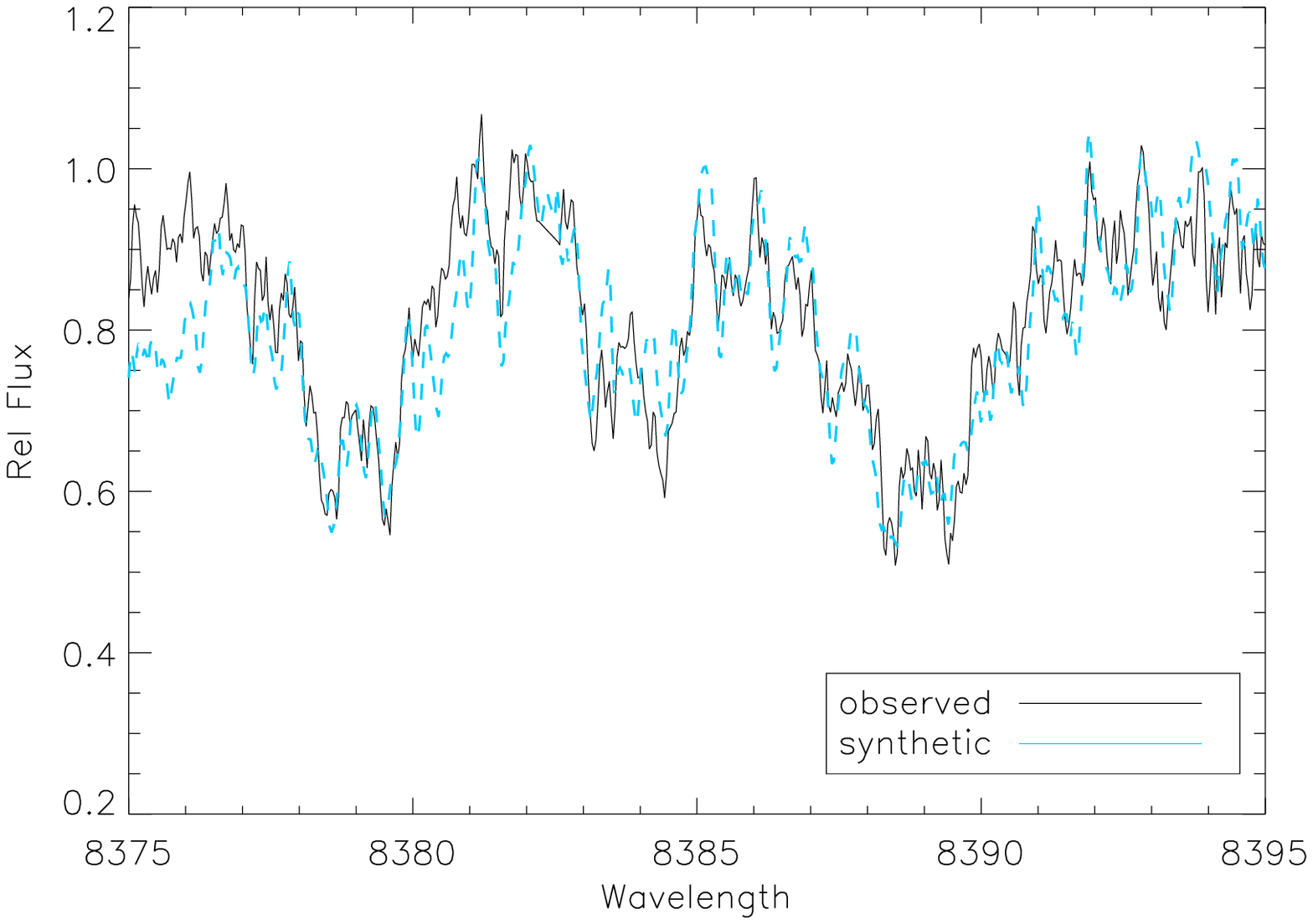}
      \caption{Spectra of 2MASS123-3610 in two different regions, 
 centred in Na~{\sc i} 8183 \AA \ on the left and in 8385 \AA \ on the right. 
 The observed spectrum is plotted as a solid 
 line and the synthetic spectra (made from suitable templates) as a dashed line.
}
         \label{fig:fig3b}
   \end{figure*}

 In order to obtain an estimate of the spectral type 
 we use other objects observed in the same run to  
 construct synthetic spectra by using the program {\sc starmod} (see Sect. 4.4).
{\sc starmod} allows to rotationally broaden spectra and shift
them in the velocity space. We ran it on the spectra of several spectral 
 types objects and combined with the appropriate weights to create
 composite spectra that were used as template for comparison with
 our target. 
 The best fit we find corresponds to a M7 primary
  and at least M7 or later secondary in a 50/50\% contribution,
 but earlier primary is possible and more spectra would be 
 necessary to improve the results with this technique. The two wavelength regions
 chosen for the fits are plotted in Fig.~\ref{fig:fig3b}, centred in  
 the first line of Na~{\sc i} doublet ($\lambda\lambda$ 8183, 8195 \AA) 
 and centred in  $\approx$8385 \AA \ where there are several Ca~{\sc i} 
 and Fe~{\sc i} lines.
 The observed spectra of 2MASS0123-3610 is plotted as continuous line
 and the correspondent synthetic spectra is overplotted as dashed line.

\begin{figure*}
   \centering
   \includegraphics[width=6.0cm,angle=90,clip=]{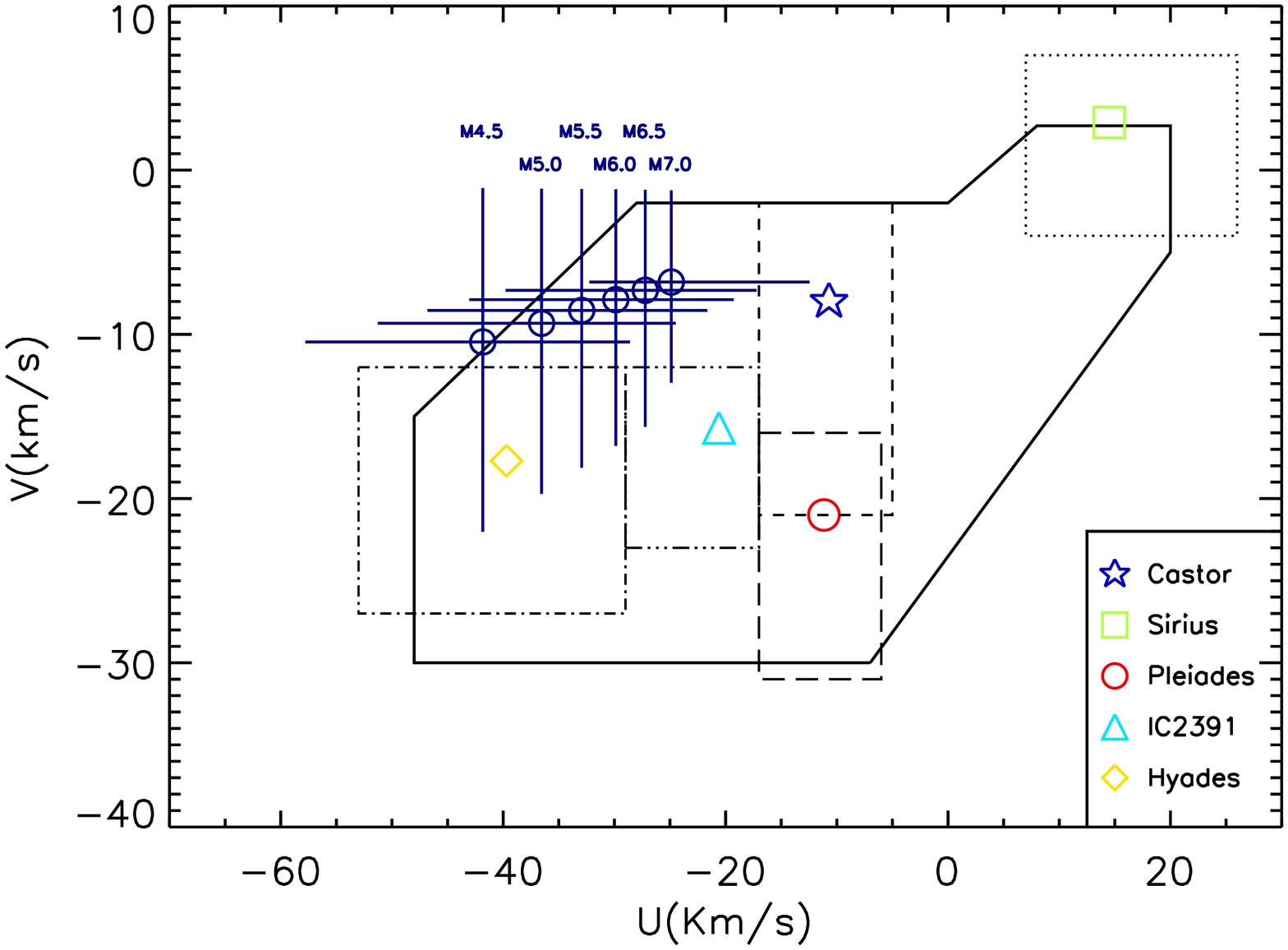}
   \includegraphics[width=6.0cm,angle=90,clip=]{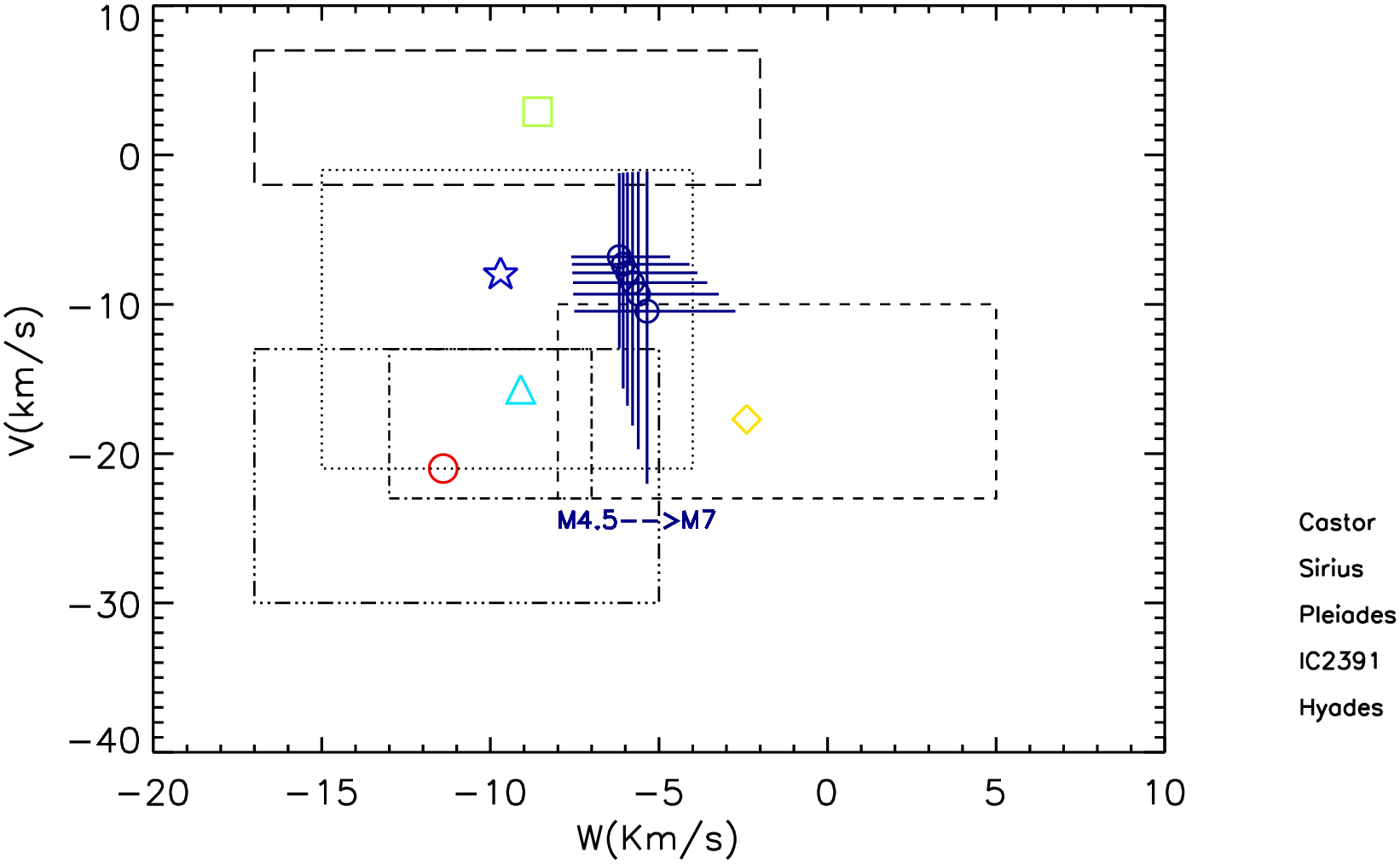}

      \caption{
 We plot here the UV and VW space motion diagrams for 2MASS0123-3610.
 The crosses represent the different positions of the system
 for different (U,V,W) values obtained from the distances derived
 from spectral type information. All possible positions lie in the YD
 area and also within the Hyades or IC 2391 boxes. 
}
         \label{fig:fig4}
   \end{figure*}

From the different spectral types possible for the primary, we derive
 the corresponding distances and galactic velocity components.
 Fig.~\ref{fig:fig4} represent the correspondent position in the
 YD area in the UV and VW planes. In any case the binary lies inside the YD
 area and also inside Hyades MG or IC 2391 MG boxes.
 Taking into account these possible memberships, we
 estimate the masses of 0.15M$_{\odot}$ and 0.09M$_{\odot}$ for primary and secondary components
 respectively if the system belongs to the Hyades MG and 0.06M$_{\odot}$ and 0.04M$_{\odot}$
  if it belongs to the IC 2391 MG.

 Rotational velocity measurements give a value at the limit of our
  sensitivity ($\approx$10 km s$^{-1}$) for both components. 
 This fact, the binarity and the uncertainty of the spectral
 classification gives us no clear conclusion about its youth using
 rotational velocity criterion. 

We will take high resolution spectra to follow up the target and 
 measure radial velocities through the period in order to 
 get the orbital parameters. Due to the SB2 nature we will be able to
 get the masses of both components. 
If the system is confirmed to belong to Hyades or IC 2391, we will have a 
 presumably coeval UCD system with known age, metallicity and stellar parameters.
Identifying such objects is crucial to increase the sample 
 of benchmarks that can provide empirical constraints for star formation theories.

\section{Conclusions and Future work}
\subsection{Conclusions}
   We studied the spectra of a 68 target sample of low-mass object previously selected
 via photometric and astrometric criteria, as possible members to five known young
 moving group.
 Using high resolution spectra we measure spectral types and derive distances to
 determine galactic space-velocity components. After applying kinematic
 criterion we find that 49 targets belong to the young disk area and that
   36 of them possibly belong to one of our five moving groups.
We measure when possible projected rotational velocities in order to use
 the rotation rate as a supporting criterion of youth and so confirm
 the kinematic members. We find that from the young disk targets, 
 31 have rotational velocities in agreement with their youth and 12 can not
 be dismissed as YD due to rotational velocities and
 further criteria should be applied.

These new candidates significantly increase the known
population of low-mass young MG members, which gives rise to an important
statistical improvement in calculations of each clusters IMF.

 Within the sample, we find a SB2 binary possible member of Hyades or IC 2391
 MGs. The spectral type classification gives as a $\approx$M5-M7 primary 
 and a possible later than M7 secondary. 
 The rotational velocity criterion in this case is not 
 conclusive so further criteria should be applied. Follow up spectroscopy
 will be carried out in order to improve the classification and to obtain
 orbital solution and derive physical parameters. If confirmed as a MG member,
 this object can serve as a very valuable testbed for evolutionary models.
  
\subsection{Future work}
 Further reliable spectroscopic criteria should be applied to confirm the youth of
 the candidates found here. The analysis of some age sensitive
 spectroscopic signatures such as Li~{\sc i} 6708 \AA \ presence an equivalent width,
 H$\alpha$ emission (e.g. West et al. 2008) and gravity sensitive
 features (e.g. Gorlova et al. 2003; McGovern et al. 2004) will
 provide reliable age constraints for our sample.
 As we mentioned in Section 3, UVES observing runs include Li~{\sc i} 
 but it is on a very low signal-to-noise region and so no conclusion could be reached.
  FEROS run included both H$\alpha$ and Li~{\sc i} region, but although we can see
  the emission in H$\alpha$ in nearly all FEROS targets, the signal-to-noise 
 did not allow us to do any equivalent width measurement, and the Li~{\sc i} region is 
 also unusable due to low signal. We are so currently carrying out new spectroscopic observations to 
 study these two important age features.
  Detailed information of the criteria and how we will use them 
 depending on the spectral type and MG candidature of our
 sample can be seen in Sect. 4.4.1 and Fig. 9 in Paper I.

 To improve the space motion determination and so improve the kinematic 
 criterion, we are also carrying out a parallax program that will allow us
  to measure more accurate distance.

 Discovering planetary systems around young and low-mass objects would be very 
 revealing for the understanding of planet and stellar formation. By
 studying these systems we will probe the host mass impact on the formation
 of different kind of planets and test actual theoretical formation models
 (Laughlin et al. 2004, Kornet \& Wolf 2006, Kornet et al. 2006)

 Our compiled sample of young MG members will be targeted in the search for 
 lower mass companions by imaging techniques. Their 
 proximity allows the exploration of the faint circumstellar 
 environment at relative small distances from the star and,
  since substellar companions cool and fade with time, targeting young systems
 will give higher probability of detection compared to typical older 
 field objects. 
  From the complete sample, our youngest closest MG members 
 (in the IC 2391 or Pleiades MGs) would allow AO to probe to 1M$_{\rm{J}}$
 at separations of $\sim$1AU using for example NACO (Nasmyth Adaptive 
 Optics System (NAOS), Near-Infrared Imager and Spectrograph (CONICA))
 at the VLT. 



\begin{table*}
\caption[]{Observing runs
\label{tab:obs}}
\begin{flushleft}
\begin{center}
\small
\begin{tabular}{cccccccc}
\noalign{\smallskip}
\hline  
\noalign{\smallskip}
Number & Date & Telescope & Instrument & Spect. range & Orders &
Dispersion & FWHM$^{1}$  \\
  &  &           &              & (\AA)   &          & (\AA) & (\AA) \\
\noalign{\smallskip}
\hline
\noalign{\smallskip}

1 & 28-30/12/06 \& 16-17/06/07 & ESO-2.2m & FEROS &  3500-9200 & 39 &0.033 - 0.081  & 0.076 - 0.131\\
2 & 9/10/-16/12/07 & ESO-VLT-U2 & UVES  &  6650-10425 & 33 & 0.027-0.041& 0.15-0.23 \\
3 & 23/02-25/02/08 & GEMINI & Phoenix &  15532-15606 &1 & 0.074 & - \\
4 & 28/03-21/06/08 & ESO-VLT-U2 & UVES   & 6650-10425 & 33 & 0.027-0.041& 0.15-0.23 \\
\noalign{\smallskip}
\hline
\noalign{\smallskip}
\end{tabular}
\end{center}
$^{1}$ Full Width at Half Maximum of the arc comparison lines\\
\end{flushleft}
\end{table*}

\begin{table*}
\caption[]{Stellar coordinades, astrometry, spectral types from literature are given with derived spectral types and distances
\label{tab:par}}
\begin{flushleft}
\scriptsize
\begin{center}
\begin{tabular}{lllccccccc}
\noalign{\smallskip}
\hline
\noalign{\smallskip}
      Name     &         $\alpha$ (2000)     &        $\delta$ (2000)      &   $\mu$$_{\alpha}$ cos $\delta$  &    $\mu$$_{\delta}$  &  PC3 &  SpT  & SpT (lit)  &  Distance & Distance (lit) \\
 & (h m s) & ($^{\rm o}$ ' '') & (mas yr$^{-1}$) & (mas yr$^{-1}$) & & & &  (pc) & (pc)\\
\noalign{\smallskip}
\hline
\noalign{\smallskip}
SIPS0004-5721  &     0  4 18.970  &  -57 21 23.30  &    157   &     -15   &  1.66 &   M7.0           &                 & 45.4$\pm$5.5  & \\
SIPS0007-2458  &     0  7  7.800  &  -24 58  3.80  &    192   &     -59   &  1.70 &   M7.0           &                 & 28.9$\pm$3.5  & \\
2MASS0020-2346 &     0 20 23.155  &  -23 46  5.38  &    340   &    -65   &  1.43 &    M6.0           &                 & 26.4$\pm$3.2  & \\
DENIS0021-4244 &     0 21  5.896  &  -42 44 43.33  &    262   &     -17   &  2.39 &    M9.5           &   M9.5$^{1}$    & 17.7$\pm$2.2  & \\
SIPS0027-5401  &     0 27 23.240  &  -54  1 46.20  &    429   &    -18   &  1.47 &   M6.5           &   M7.0$^{2}$    &  25.6$\pm$3.1  &  28$\pm$2.5 (Spt)$^{2}$\\
SIPS0039-2256  &     0 39 23.250  &  -22 56 44.90  &    224   &      44   &  1.74 &   M7.0           &                 & 47.9$\pm$5.8  & \\
DENIS0041-5621 &     0 41 35.390  &  -56 21 12.77  &     92   &     -62   &  - &   M7.5$^{3}$        &   M7.5$^{4}$   & 16.1$\pm$2.0  & \\
SIPS0054-4142  &     0 54 35.300  &  -41 42  6.20  &    64   &    -108   &  1.25 &   M5.0           &                &  37.9$\pm$4.6  & \\
SIPS0109-0343  &     1  9 51.040  &   -3 43 26.30  &    207   &     87   &  2.06 &   M9.0          &   M9.5$^{5}$   &  11.0$\pm$1.3  & 9.59$\pm$0.2 (plx)$^{5}$\\
LEHPM1289      &     1  9 59.579  &  -24 16 47.82  &    366   &      -4   &  1.47 &   M6.5           &                 &  45.2$\pm$5.5  & \\
SIPS0115-2715  &     1 15 26.610  &  -27 15 54.10  &    149   &      32   &  1.27 &   M5.5           &                 &  87.7$\pm$10.7  & \\
2MASS0123-3610* &     1 23  0.506  &  -36 10 30.67  &    124   &      57   & 1.19 &    M5.0           &                &  66.1$\pm$8.1  & \\
SIPS0126-1946  &     1 26 49.980  &  -19 46  5.90  &    210   &     -11   &  1.45 &   M6.5           &                 & 62.3$\pm$7.6  & \\
LEHPM1563      &     1 27 31.956  &  -31 40  3.18  &    290   &     131   &  1.95 &   M7.0          &                 & 18.9$\pm$2.3  & \\
SIPS0153-5122  &     1 53 11.430  &  -51 22 24.99  &    127   &      27   &  1.43 &   M6.0           &                 & 44.1$\pm$5.4  & \\
2MASS0204-3945 &     2  4 18.036  &  -39 45  6.48  &    119   &      16   &  1.66 &   M7.0           &                & 33.9$\pm$4.1  & \\
SIPS0212-6049  &     2 12 33.580  &  -60 49 18.40  &    112   &     -25   &  1.55 &   M6.5           &                 & 36.3$\pm$4.4  & \\
SIPS0214-3237  &     2 14 45.440  &  -32 37 58.20  &    148   &      42   &  1.49 &   M6.0           &                 & 53.2$\pm$6.5  & \\
SIPS0235-0711  &     2 35 49.470  &   -7 11 21.90  &    286   &     75   &  1.41 &   M6.0           &   M5.5$^{4}$    & 28.2$\pm$3.4  & \\
2MASS0334-2130 &     3 34 10.657  &  -21 30 34.35  &    124   &       7   &  - &   M4.5$^{3}$     &   M6.0$^{6}$   &  13.6$\pm$4.1  & 23.3 $\pm$4.0(Spt)$^{6}$\\
2MASS0429-3123 &     4 29 18.426  &  -31 23 56.81  &    106   &      69   &  - &   M7.0$^{3}$     & M7.5-L0$^{7}$ (binary)  & 9.7$\pm$0.9  & \\
SIPS0440-0530  &     4 40 23.328  &   -5 30  7.85  &    330   &     133   &   1.74 &  M7.0          &  M7.5$^{2}$     &  9.0$\pm$1.1  & \\
2MASS0445-5321 &     4 45 43.368  &  -53 21 34.56  &   343    &    485    &   -  &  M7.5$^{phot}$  &        &  25.5$\pm$1.3  & \\
2MASS0502-3227 &     5  2 38.677  &  -32 27 50.07  &    063   &    -172   &   1.37 &  M6          &  M6.0$^{6}$      &  29.6$\pm$3.6  & 25.1 $\pm$ 3.4(Spt)$^{6}$\\
2MASS0528-5919 &     5 28  5.623  &  -59 19 47.17  &     77   &     152   &   1.34 &  M6           &                   & 82.0$\pm$10.0  &  \\
2MASS0600-3314 &     6  0 33.750  &  -33 14 26.84  &   -28    &    153    &   -  & M7.0$^{phot}$   &  M7.5$^{6}$      & 33.4$\pm$2.6  & 28.0 $\pm$ 2.7(Spt)$^{6}$\\
 SIPS1039-4110 &    10 39 18.340  &  -41 10 32.00  &    25    &   -173    &  - &  M6.5$^{phot}$   &  M6.0$^{8}$      &  20.8$\pm$5.3 & \\
 SIPS1124-2019 &    11 24 22.229  &  -20 19  1.50  &   129    &    -19    &  - &  M7.0$^{phot}$   &                  &  37.5$\pm$2.5 & \\
DENIS1250-2121 &    12 50 52.654  &  -21 21 13.67  &    457   &    -320   &   - &  M7.5$^{3}$         &  M7.5$^{2}$      & 9.8$\pm$0.4  &  11.1 $\pm$ 1.3(Spt)$^{2}$\\
 SIPS1329-4147 &    13 29  0.872  &  -41 47 11.90  &   293    &   -302    &  - &  M9.5$^{phot}$   &  M9.0$^{8}$      & 27.3$\pm$2.4 & \\
 SIPS1341-3052 &    13 41 11.561  &  -30 52 49.60  &   109    &   -163    &  - &  L0$^{phot}$     &                  & 40.8$\pm$11.5  & \\
2MASS1507-2000 &    15  7 27.799  &  -20  0 43.18  &    114    &    -76   &  - &   M7.5$^{3}$      &  M7.5$^{6}$      & 15.2$\pm$0.8  &  11.0 $\pm$ 0.2(Spt)$^{6}$\\
SIPS1632-0631  &    16 32 58.799  &   -6 31 45.30  &    29   &    -366   &  1.95 &   M8.5          &  M7.0$^{8}$     &  19.7$\pm$2.4  & \\
SIPS1758-6811  &    17 58 59.663  &  -68 11 10.50  &    3   &    -182   &   1.25 &  M5.5          &                 & 71.4$\pm$8.7  &\ \\
SIPS1949-7136  &    19 49 45.527  &  -71 36 50.89  &    36   &    -183   &    1.59 & M7.0           &                 & 46.2$\pm$5.6  & \\
SIPS2000-7523  &    20  0 48.171  &  -75 23  6.58  &    179   &    -85   &  1.88 &   M8.0           &                 & 20.9$\pm$2.5  & \\
2MASS2001-5949 &    20  1 24.639  &  -59 49  0.09  &    163   &    -54   &  1.46 &   M6.0           &                 & 38.6$\pm$4.7  & \\
SIPS2014-2016  &    20 14  3.523  &  -20 16 21.30  &    248   &    -112   &  1.71 &   M7.5          &  M7.5$^{6}$     & 22.0$\pm$2.7  & \\
2MASS2031-5041 &    20 31 27.495  &  -50 41 13.49  &    161   &    -160   &   1.31 &  M5.5           &                 & 48.3$\pm$5.9  & \\
SIPS2039-1126  &    20 39 13.081  &  -11 26 52.30  &    64   &    -105   &  1.61 &   M7.0           &  M8.0$^{6}$     & 42.6$\pm$5.2  & \\
SIPS2045-6332  &    20 45  2.278  &  -63 32  5.30  &    97   &    -201   &   2.16 &  M8.5           &                 & 15.3$\pm$1.9  & \\
SIPS2049-1716  &    20 49 52.610  &  -17 16  7.80  &    369   &    -142   &   - &  M6.5$^{3}$         &  M6.0$^{9}$     & 13.2$\pm$1.6  & \\
SIPS2049-1944  &    20 49 19.673  &  -19 44 31.30  &    179   &    -279   &   1.60 &  M7.0           &  M7.5$^{10}$     & 28.1$\pm$3.4  & \\
SIPS2100-6255  &    21  0 30.227  &  -62 55  7.31  &     65   &    -110   &  1.19 & M5.0 & M5.0$^{phot}$      & 66.1$\pm$10.2  & \\
2MASS2106-4044 &    21  6 20.896  &  -40 44 51.91  &    150   &    -83   &  1.47 &   M6.0           &                 & 38.8$\pm$4.7  & \\
SIPS2114-4339  &    21 14 40.928  &  -43 39 51.20  &    49   &    -148   &   1.57 &  M6.5           &                 & 31.1$\pm$3.8  & \\
SIPS2119-0740  &    21 19 17.571  &   -7 40 52.50  &    152   &    -117   &   1.64 &  M7.0           &                 & 47.0$\pm$5.7  & \\
HB2124-4228    &    21 27 26.133  &  -42 15 18.39  &    97   &    -170   &   1.76 &  M8.0           &  M8.5$^{11}$   & 29.9$\pm$3.6  & 35.7$\pm$10.1(plx)$^{12}$\\
SIPS2128-3254  &    21 28 17.402  &  -32 54  3.90  &    338   &    -166   &   - &  M6.5$^{3}$       &            &  33.0$\pm$1.0  & \\
DENIS2200-3038 &    22  0  2.022  &  -30 38 32.71  &    199   &    -66   &  2.07 &   M9.0        &  M9-L0$^{13}$   & 24.8$\pm$3.0  & 35 $\pm$ 2(Spt$^{13}$)\\
SIPS2200-2756  &    22  0 16.838  &  -27 56 29.70  &    156   &    -8   &   1.46 &  M6.0           &                &  35.5$\pm$4.4  &\\
2MASS2207-6917 &    22  7 10.313  &  -69 17 14.25  &    131   &    -51   &   1.49 &  M6.5           &                &  43.2$\pm$5.3  &\\
LEHPM4480      &    22 15 10.151  &  -67 38 49.07  &    260   &    -136   &  1.33 &   M6.0           &                & 51.9$\pm$6.3  & \\
2MASS2222-4919 &    22 22  3.684  &  -49 19 23.45  &    67   &    -122   &  1.38 &   M6.0           &                & 76.9$\pm$9.4  & \\
2MASS2231-4443 &    22 31  8.657  &  -44 43 18.43  &    163   &    -9   &   1.23 &  M5.0           &                & 46.3$\pm$5.6  & \\
LEHPM4908      &    22 36 42.656  &  -69 34 59.30  &    220   &    -63   &   1.39 &  M6.0           &                & 32.3$\pm$3.9  & \\
2MASS2242-2659 &    22 42 41.294  &  -26 59 27.23  &    99   &    -26   &   1.33 &  M5.5           &                & 47.4$\pm$5.8  & \\
2MASS2254-3228 &    22 54 58.110  &  -32 28 52.20  &    55   &    -84   & 1.37 &    M6.0           &                & 49.7$\pm$6.1  & \\
2MASS2311-5256 &    23 11 30.330  &  -52 56 30.17  &    141   &     -43   &   1.36 &  M5.5           &                & 66.5$\pm$8.1  & \\
SIPS2318-4919  &    23 18 45.952  &  -49 19 17.79  &    227   &     -25   &   - &  M7.0$^{phot}$      &   M8.0$^{2}$             & 48.9$\pm$3.7  & \\
SIPS2321-6106  &    23 21 43.418  &  -61  6 35.37  &    110   &      58   &   1.31 &  M5.5           &                & 49.5$\pm$6.0  & \\
SIPS2322-6357  &    23 22  5.332  &  -63 57 57.60  &    122   &     -28   &  1.54 &   M6.5           &  M7.0$^{2}$    & 56.4$\pm$6.9  & \\
SIPS2341-3550  &    23 41 47.497  &  -35 50 14.40  &    154   &     -28   &  1.64 &   M7.0           &                & 36.7$\pm$4.5  & \\
SIPS2343-2947  &    23 43 34.731  &  -29 47  9.50  &    257   &      -8   &  1.83 &   M8.0           &                & 32.3$\pm$3.9  & \\
SIPS2347-1821  &    23 47 16.662  &  -18 21 50.60  &    219   &      41   &  1.57 &   M6.5           &                & 31.8$\pm$3.9  & \\
SIPS2350-6915  &    23 50  3.948  &  -69 15 24.39  &    164   &      11   &  1.52 &   M6.0          &                & 56.1$\pm$6.8  & \\
LEHPM6375      &    23 52 49.138  &  -22 49 29.54  &    222   &    -175   &   1.51 &  M6.0           &                & 33.0$\pm$4.0  & \\
LEHPM6542      &    23 57 54.822  &  -19 55  1.89  &    185   &      -4   &   1.41 &  M6.0           &                & 42.3$\pm$5.2  & \\
\noalign{\smallskip}
\hline
\noalign{\smallskip}
\end{tabular}
\end{center}
$^{1}$ Basri et al. 2000, ApJ, 538, 363. $^{2}$ Lodieu et al. 2005, A\&A 440, 1061. 
$^{3}$ Feros run object. See Sect 4.1 and Paper I. $^{4}$ Phan-Bao \& Bessel, 2006, A\&A, 446, 51. 
 $^{5}$ Costa et al, 2005, AJ, 130, 337. $^{6}$ Cruz et al. 2003, AJ, 126, 2421. $^{7}$ Siegler et al., 2005,621, 1023.
 $^{8}$ Gizis, 2002, ApJ, 575, 484. $^{9}$ Crifo et al., 2005, A\&A 441, 653. $^{10}$ Basri \& Reiners, 2006, AJ, 132, 663.
 $^{11}$ Burgasser et al., 2002, ApJ, 564, 421. $^{12}$ Tinney, 1996, MNRAS, 281, 644. $^{13}$ Burgasser, 2006, AJ, 131, 1007. 
$^{phot}$ Spectral types derived by J-I colors. See text. '*' See Sect. 5.3 \\
\end{flushleft}
\end{table*}

\begin{table*}
\caption[Reference stars]{Radial and Rotational are given for our reference stars
\label{tab:ref}}
\begin{flushleft}
\scriptsize
\begin{center}
\begin{tabular}{lcccccc}
\hline
\noalign{\smallskip}
Name & SpT & $V_{\rm r}$& Ref$_{\rm V_r}$ & $V\sin{i}$  & Ref$_{\rm V\sin{i}}$  & Observing  \\
   &              & (km s$^{-1}$) &    &(km s$^{-1}$) &    &   run      \\
\noalign{\smallskip}
\hline
\noalign{\smallskip}
GI 876 & M4.0 & -1.591 & $^{1}$  & $<$2.0 & $^{2}$ & 1,2,4 \\
GL 406 & M6.0 & -19.402 & $^{1}$ & $<$3.0 & $^{2}$ & 1,3\\
VB 10 & M8.0 & 34.7 & $^{3}$ & 6.5 & $^{2}$ &  1,4 \\
GJ 3877 & M7.0 & 1.4 & $^{2}$ & 8.0 & $^{2}$ & 1  \\
LP 944-20 & M9.0 & 10 & $^{4}$ & 30.3 &$^{5}$&  1,2,3\\
\noalign{\smallskip}
\hline
\noalign{\smallskip}
\end{tabular}
\end{center}
$^{1}$ Nidever et al., 2002, ApJ, 141, 503\\
$^{2}$ Mohanty, 2003, ApJ, 583, 451\\
$^{3}$ Mart\'in et al., 2006, ApJ, 644, L75\\
$^{4}$ Ribas 2003, A\&A, 398, 239R\\
$^{5}$ Zapatero-Osorio et al, 2006, AJ, 647, 1405\\
\end{flushleft}
\end{table*}

\begin{table*}
\caption[]{MG assignment are given based on galactic velocity coordinates. (-) indicates values not measured due to the low signal-to-noise of the spectra.
\label{tab:vrs}}
\begin{flushleft}
\scriptsize
\begin{center}
\begin{tabular}{lccrrrcccc}
\noalign{\smallskip}
\hline 
\noalign{\smallskip}
Name        &  $V_{\rm r}$ $\pm$ $\sigma_{V_{\rm r}}$ & distance & $U \pm \sigma_{\rm U}$  &  $V \pm \sigma_{\rm V}$ &  $W \pm \sigma_{\rm W}$&  MG & $V\sin{i}$  & Observing & S/N \\
 & (km s$^{-1}$) & (pc) & (km s$^{-1}$) & (km s$^{-1}$) & (km s$^{-1}$) & candidature & (km s$^{-1}$) & run & \\
\noalign{\smallskip}
\hline
\noalign{\smallskip}
SIPS0004-5721  &  6.54$\pm$0.46 & 69.3 $\pm$ 8.5  &  -41.0 $^{+  12.3}_{-   14.2}$    &  -29.8 $^{+  10.7 }_{-   12.8 }$   &  -12.6 $^{+  5.1}_{-  6.1}$    &  {\bf IC 2391} & 20.11 & 2 & 22\\ 
 ''             &  & 45.4 $\pm$ 5.5  &  -26.0 $^{+  8.1}_{-   9.3}$    &  -20.3 $^{+  7.1}_{-   8.4}$    &  -10.1 $^{+  3.5}_{-   4.1}$   & {\bf Hyades}  & & & \\
''             &  & 51.6 $\pm$ 6.3  &  -29.9 $^{+  9.2 }_{-   10.6 }$    &  -22.8 $^{+  8.0 }_{-   9.5 }$    &  -10.8 $^{+  3.9 }_{-   4.7 }$    &  Castor & & &  \\
SIPS0007-2458 & 10.43$\pm$0.46  &  38.0 $\pm$ 4.6  &  -23.7 $^{+  5.1 }_{-   5.6 }$    &  -24.0 $^{+  5.1 }_{-   5.6 }$    &  -16.7 $^{+  1.5 }_{-   1.6 }$    &  Pleiades  & 17.96  & 2 & 24\\
''         &    &  44.2 $\pm$ 5.4  &  -27.9 $^{+  5.9 }_{-   6.5 }$    &  -28.1 $^{+  5.9 }_{-   6.5 }$    &  -17.7 $^{+  1.7 }_{-   1.8 }$    &  {\bf IC 2391}  & & & \\
''         &    &  32.9 $\pm$ 4.0 &  -20.4 $^{+  4.4 }_{-   4.9 }$    &  -20.6 $^{+  4.4 }_{-   4.9 }$    &  -15.8 $^{+  1.4 }_{-   1.5 }$    &  Castor  & & & \\
2MASS0020-2346  & 2.19$\pm$0.15 & 40.3 $\pm$ 4.9  &  -49.9 $^{+  9.4 }_{-   10.3 }$    &  -42.2 $^{+  8.5 }_{-   9.4 }$    &  -10.8 $^{+  1.6 }_{-   1.7 }$    &  IC 2391 & 12.32 & 4 & 24\\
DENIS0021-4244  & 2.00$\pm$3.00 &  27.1 $\pm$ 3.3  &  -27.1 $^{+  7.0 }_{-   8.0 }$    &  -19.0 $^{+  5.8 }_{-   6.8 }$    &  -7.5 $^{+  1.8 }_{-   2.1 }$    &  {\bf IC 2391} & 11.39 & 2 & 20\\
 ''          &   &  17.7 $\pm$ 2.2  &  -17.4 $^{+  4.6 }_{-   5.3 }$    &  -12.7 $^{+  3.8 }_{-   4.5 }$    &  -6.3 $^{+  1.3 }_{-   1.5 }$    &  Hyades  & & &\\
''           &  &  20.2 $\pm$ 2.5  &  -19.9 $^{+  5.2 }_{-   6.0 }$    &  -14.4 $^{+  4.3 }_{-   5.1 }$    &  -6.6 $^{+  1.4 }_{-   1.6 }$    &  {\bf Castor}  & & &\\
SIPS0027-5401  &  -11.49$\pm$0.88 & 39.1 $\pm$ 4.8  &  -70.5 $^{+  13.3 }_{-   14.6 }$    &  -38.4 $^{+  10.4 }_{-   11.7 }$    &  4.3 $^{+  3.6 }_{-   4.1 }$    &  IC 2391  & 27.43 & 4 & 22\\
 ''          &   &  25.6 $\pm$ 3.1  &  -47.3 $^{+  8.8 }_{-   9.7 }$    &  -23.8 $^{+  6.9 }_{-   7.8 }$    &  6.3 $^{+  2.6 }_{-   3.0 }$    &  {\bf Hyades}  & & &\\
SIPS0039-2256 & -16.04$\pm$0.26 &  47.9 $\pm$ 5.8  &  -48.6 $^{+  10.7 }_{-   12.0 }$    &  -19.5 $^{+  7.0 }_{-   8.3 }$    &  14.3 $^{+  0.9 }_{-   1.0 }$    &  Hyades  & 12.82 & 2 & 20\\
DENIS0041-5621  &  2.36$\pm$1.03 & 18.3 $\pm$ 2.2  &  -3.7 $^{+  2.3 }_{-   2.7 }$    &  -9.2 $^{+  2.9 }_{-   3.3 }$    &  0.3 $^{+  2.0 }_{-   1.8 }$    &  Sirius  & $\approx$15$^{1}$ & 1 & 21\\
''           &   &  21.1 $\pm$ 2.6  &  -4.4 $^{+  2.6 }_{-   3.0 }$    &  -10.4 $^{+  3.3 }_{-   3.7 }$    &  0.6 $^{+  2.1 }_{-   1.9 }$    &  Pleiades  &  & &\\
 ''          &   &  18.3 $\pm$ 2.2  &  -3.7 $^{+  2.3 }_{-   2.7 }$    &  -9.2 $^{+  2.9 }_{-   3.3 }$    &  0.3 $^{+  2.0 }_{-   1.8 }$    &  Castor  & & &\\
SIPS0054-4142 & -5.81$\pm$0.25  &  43.1 $\pm$ 5.3  &  0.1 $^{+  3.5 }_{-   3.3 }$    &  -23.7 $^{+  5.7 }_{-   6.4 }$    &  11.3 $^{+  1.6 }_{-   1.4 }$    &  Castor & $<$10 & 4 & 26\\
LEHPM1289  & 23.36$\pm$0.10 & 45.2 $\pm$ 5.5  &  -65.6 $^{+  14.3 }_{-   16.1 }$    &  -45.6 $^{+  12.2 }_{-   14.0 }$    &  -17.7 $^{+  1.5 }_{-   1.4 }$    &  Hyades   & 12.62  & 2 & 26\\
SIPS0109-0343  & -10.36$\pm$0.64 & 9.6 $\pm$ 0.2  &  -6.8 $^{+  3.5 }_{-   3.6 }$    &  -5.3 $^{+  3.4 }_{-   3.5 }$    &  11.7 $^{+  1.8 }_{-   1.7 }$    &  Hyades  & & 4 & 27\\
SIPS0115-2715  & 43.51$\pm$0.12 & 87.7 $\pm$ 10.7  &  -61.1 $^{+  11.4 }_{-   12.6 }$    &  -27.5 $^{+  7.4 }_{-   8.6 }$    &  -37.6 $^{+  1.2 }_{-   1.1 }$    &  Hyades  & $<$10 & 2 & 20
\\
2MASS0123-3610*  &  7.53$\pm$0.29* & 101.0 $\pm$ 12.3  &  -63.9 $^{+  20.5 }_{-   24.0 }$    &  -15.2 $^{+  14.3 }_{-   17.6 }$    &  -4.3 $^{+  3.8 }_{-   3.1 }$    &  IC 2391  & $<$10/$<$10 & 2 & 24\\
''            &  &  66.1 $\pm$ 8.1  &  -41.8 $^{+  13.4 }_{-   15.7 }$    &  -10.5 $^{+  9.4 }_{-   11.6 }$    &  -5.4 $^{+  2.6 }_{-   2.2 }$    &  {\bf Hyades} &  & & \\
SIPS0126-1946  & 28.89$\pm$0.14 & 62.23 $\pm$ 7.6  &  -52.6 $^{+  17.6 }_{-   20.9 }$    &  -38.9 $^{+  16.9 }_{-   20.2 }$    &  -20.3 $^{+  4.1 }_{-   3.4 }$    &  Hyades  & $>$13$^{1}$ &2 & 19\\
LEHPM1563  &  27$\pm$0.25 & 18.9 $\pm$ 2.3  &  -29.5 $^{+  5.0 }_{-   5.5 }$    &  -10.1 $^{+  2.5 }_{-   2.9 }$    &  -23.9 $^{+  0.9 }_{-   0.8 }$    &  Hyades  & 15.25  & 2 & 27\\
SIPS0153-5122 & 1.35$\pm$0.24 &  57.8 $\pm$ 7.1  &  -29.6 $^{+  7.8 }_{-   9.0 }$    &  -23.0 $^{+  5.8 }_{-   6.9 }$    &  -6.5 $^{+  3.2 }_{-   2.7 }$    &  Pleiades  & 13.67 & 2 & 22\\
''         &    &  67.3 $\pm$ 8.2  &  -34.6 $^{+  9.1 }_{-   10.4 }$    &  -25.8 $^{+  6.8 }_{-   8.0 }$    &  -5.6 $^{+  3.7 }_{-   3.1 }$    &  {\bf IC 2391}   & & & \\
''         &    &  44.1 $\pm$ 5.4  &  -22.3 $^{+  6.0 }_{-   6.8 }$    &  -18.9 $^{+  4.5 }_{-   5.3 }$    &  -7.8 $^{+  2.5 }_{-   2.1 }$    &  {\bf Hyades}   & & &\\
2MASS0204-3945  & 11.90$\pm$0.13 & 44.5 $\pm$ 5.4  &  -21.3 $^{+  6.1 }_{-   7.1 }$    &  -17.5 $^{+  5.1 }_{-   6.0 }$    &  -4.9 $^{+  2.5 }_{-   2.1 }$    &  {\bf Pleiades} &  $<$10  & 2 & 22\\
''           &   &  51.8 $\pm$ 6.3  &  -24.6 $^{+  7.1 }_{-   8.3 }$    &  -19.8 $^{+  5.9 }_{-   7.0 }$    &  -3.8 $^{+  2.9 }_{-   2.5 }$    &  {\bf IC 2391}  & & &\\
''           &   &  33.9 $\pm$ 4.1  &  -16.4 $^{+  4.7 }_{-   5.4 }$    &  -14.3 $^{+  3.9 }_{-   4.6 }$    &  -6.4 $^{+  1.9 }_{-   1.7 }$    &  Hyades  & & &\\
''           &   &  38.6 $\pm$ 4.7  &  -18.6 $^{+  5.3 }_{-   6.2 }$    &  -15.7 $^{+  4.4 }_{-   5.2 }$    &  -5.7 $^{+  2.2 }_{-   1.9 }$    &  {\bf Castor}  & & &\\
SIPS0212-6049 & 13.11$\pm$0.15 &  41.3 $\pm$ 5.0  &  -9.8 $^{+  3.5 }_{-   4.1 }$    &  -24.1 $^{+  3.8 }_{-   4.3 }$    &  -1.3 $^{+  2.8 }_{-   2.4 }$    &  Sirius &  $<$10  & 2 & 23\\
SIPS0214-3237  & 4.01$\pm$0.37 &  81.3 $\pm$ 9.9  &  -51.7 $^{+  17.2 }_{-   20.3 }$    &  -26.2 $^{+  13.9 }_{-   16.9 }$    &  13.2 $^{+  6.3 }_{-   5.4 }$    &  IC 2391  & 18.14 & 2 & 22\\
''           &  &  53.2 $\pm$ 6.5  &  -34.1 $^{+  11.3 }_{-   13.3 }$    &  -17.5 $^{+  9.1 }_{-   11.1 }$    &  7.3 $^{+  4.2 }_{-   3.7 }$    &  {\bf Hyades} &  & &\\
SIPS0235-0711 & 16.61$\pm$31   &  28.2 $\pm$ 3.4  &  -39.0 $^{+  5.6 }_{-   6.1 }$    &  -17.4 $^{+  4.2 }_{-   4.8 }$    &  4.4 $^{+  4.0 }_{-   3.6 }$    &  {\bf Hyades}  & 21.65 & 4 & 32 \\
2MASS0334-2130  & 18.98$\pm$0.80 & 20.7 $\pm$ 6.3  &  -15.9 $^{+  3.5 }_{-   4.5 }$    &  -13.9 $^{+  3.9 }_{-   5.0 }$    &  -7.9 $^{+  4.2 }_{-   3.5 }$    &  {\bf IC 2391} &  $<$10$^{1}$  & 1 & 26\\
2MASS0429-3123  & 41.92$\pm$0.88 &  14.8 $\pm$ 1.4  &  -25.2 $^{+  2.2 }_{-   2.5 }$    &  -26.3 $^{+  1.9 }_{-   2.2 }$    &  -22.5 $^{+  2.2 }_{-   2.0 }$    &  IC 2391  & $<$10$^{1}$ & 1 & 26\\
''           &   &  11.0 $\pm$ 1.0  &  -23.7 $^{+  1.7 }_{-   1.9 }$    &  -25.8 $^{+  1.6 }_{-   1.7 }$    &  -24.0 $^{+  1.8 }_{-   1.7 }$    &  Castor  & & &\\
SIPS0440-0530 & 28.28$\pm$0.32 & 10.2  $\pm$ 1.2  &  -29.9$^{+   1.9  2.1}$  &     -14.3$^{+   1.9}_{-   2.3}$  &      -0.2$^{+   3.3}_{-   3.0}$  &     Sirius & 14.79 & 2 & 31\\
'' & & 11.8  $\pm$ 1.4  &  -31.0$^{+   2.2}_{-   2.4}$  &     -15.1$^{+  2.2}_{-  2.6}$  &       2.0$^{+   3.8}_{-   3.4}$  &    Pleiades & & &\\
'' & & 13.7  $\pm$ 1.7  &  -32.5$^{+   2.5}_{-   2.7}$  &     -16.1$^{+   2.6}_{-   3.0}$  &       4.8$^{+   4.4}_{-   4.0}$  &      IC 2391 & & &\\
'' & & 9.0  $\pm$ 1.1   & -29.0$^{+   1.7}_{-   1.9}$  &     -13.6$^{+   1.7}_{-   2.0}$  &      -2.0$^{+   2.9}_{-   2.7}$  &      {\bf Hyades} & & &\\
'' & &  10.2  $\pm$ 1.2  &  -29.9$^{+   1.9}_{-   2.1}$  &     -14.3$^{+   1.9}_{-   2.3}$  &      -0.2$^{+   3.3}_{-   3.0}$  &      Castor & & &\\
2MASS0445-5321 & 68.46$\pm$1.14   &  25.5 $\pm$ 1.3  &  -74.1 $^{+  4.8 }_{-   5.0 }$    &  -63.4 $^{+  2.3 }_{-   2.4 }$    &  -18.4 $^{+  3.0 }_{-   2.9 }$    &  Hyades   & - & 3& 14\\
2MASS0502-3227  & -4.33$\pm$0.15 &  33.7 $\pm$ 4.1  &  24.3 $^{+  5.4 }_{-   4.8 }$    &  -15.8 $^{+  4.6 }_{-   5.2 }$    &  5.9 $^{+  3.1 }_{-   2.6 }$    &  Sirius  & 12.28 & 4 & 22\\
2MASS0528-5919  & -6.81$\pm$0.48 & 82.0 $\pm$ 10.0  &  -60.7 $^{+  15.5 }_{-   17.8 }$    &  -7.2 $^{+  6.0 }_{-   7.1 }$    &  26.3 $^{+  10.3 }_{-   8.7 }$    &  Hyades & $<$10 & 2 & 17\\
2MASS0600-3314  & 35.36$\pm$5.00 & 33.4 $\pm$ 2.6  &  -38.1 $^{+  6.9 }_{-   7.4 }$    &  -16.3 $^{+  7.9 }_{-   7.4 }$    &  -11.8 $^{+  6.3 }_{-   5.9 }$    &  {\bf Hyades}  & - & 3& 18\\
SIPS1039-4110  & 9.98$\pm$5.00 & 23.7 $\pm$ 6.0  &  13.9 $^{+  5.0 }_{-   4.5 }$    &  -11.7 $^{+  5.6 }_{-   5.7 }$    &  -12.3 $^{+  5.7 }_{-   6.1 }$    &  Sirius   & - & 3 & 15\\
SIPS1124-2019  & 13.51$\pm$5.00 & 57.3 $\pm$ 3.8  &  34.3 $^{+  10.1 }_{-   9.2 }$    &  -0.5 $^{+  9.4 }_{-   8.8 }$    &  16.2 $^{+  9.4 }_{-   8.7 }$    &  IC 2391   & - & 3 & 19\\
''          &   &  42.7 $\pm$ 2.8  &  25.9 $^{+  7.6 }_{-   7.0 }$    &  -3.0 $^{+  8.0 }_{-   7.5 }$    &  14.2 $^{+  7.8 }_{-   7.2 }$    &  Castor  & & \\
DENIS1250-2121  & -7.59$\pm$0.38 &  12.8 $\pm$ 0.6 &  27.3 $^{+  2.8 }_{-   2.7 }$    &  9.0 $^{+  1.6 }_{-   1.5 }$    &  -19.5 $^{+  1.6 }_{-   1.7 }$    &  Pleiades & - & 1 & 33\\
SIPS1329-4147  &  34.15$\pm$5.00 & 35.8 $\pm$ 3.2  &  65.3 $^{+  9.8 }_{-   9.4 }$    &  -13.3 $^{+  7.6 }_{-   7.1 }$    &  -42.6 $^{+  9.4 }_{-  10.0 }$    &  Pleiades & - & 3 &13 \\
SIPS1341-3052  & 33.68$\pm$5.00 & 46.5 $\pm$ 13.0  &  46.1 $^{+  14.5 }_{-   10.6 }$    &  -23.9 $^{+  7.0 }_{-   9.0 }$    &  -17.4 $^{+  14.8 }_{-   16.8 }$    &  Sirius  & - & 3 & 20\\
''             &  & 53.6 $\pm$ 15.0  &  50.0 $^{+  16.2 }_{-   14.1 }$    &  -24.4 $^{+  7.6 }_{-   9.9 }$    &  -22.7 $^{+  16.7 }_{-   19.0 }$    &  Pleiades  & & &\\
''             &  & 46.5 $\pm$ 13.0  &  46.1 $^{+  14.5 }_{-   12.6 }$    &  -23.9 $^{+  7.0 }_{-   9.0 }$    &  -17.4 $^{+  14.8 }_{-   16.8 }$    &  Castor  & & &\\
2MASS1507-2000  & -22.10$\pm$1.33 & 17.3 $\pm$ 0.9  &  -11.2 $^{+  2.4 }_{-   2.3 }$    &  7.8 $^{+  2.3 }_{-   2.1 }$    &  -20.8 $^{+  2.6 }_{-   2.8 }$    &  Sirius & -  & 1 & 20\\
''              &  & 19.9 $\pm$ 1.0  &  -10.1 $^{+  2.6 }_{-   2.5 }$    &  8.1 $^{+  2.6 }_{-   2.3 }$    &  -22.2 $^{+  2.9 }_{-   3.1 }$    &  Pleiades &  & &\\
''              &  & 23.2 $\pm$ 1.2  &  -8.9 $^{+  2.9 }_{-   2.8 }$    &  8.4 $^{+  2.9 }_{-   2.7 }$    &  -23.8 $^{+  3.2 }_{-   3.5 }$    &  IC 2391 &  & &\\
''              &  & 17.3 $\pm$ 0.9  &  -11.2 $^{+  2.4 }_{-   2.3 }$    &  7.8 $^{+  2.3 }_{-   2.1 }$    &  -20.8 $^{+  2.6 }_{-   2.8 }$    &  Castor  & & &\\
\noalign{\smallskip}
\hline
\noalign{\smallskip}
\end{tabular}
\end{center}
\end{flushleft}
\end{table*}
\begin{table*}
\contcaption{
\label{tab:vrs2}}
\begin{flushleft}
\scriptsize
\begin{center}
\begin{tabular}{lccrrrcccc}
\noalign{\smallskip}
\hline
\noalign{\smallskip}
Name        &  $V_{\rm r}$ $\pm$ $\sigma_{V_{\rm r}}$ & distance & $U \pm \sigma_{\rm U}$  &  $V \pm \sigma_{\rm V}$ &  $W \pm \sigma_{\rm W}$              &  MG & $V\sin{i}$ & Observing & S/N\\
 & (km s$^{-1}$) & (pc) & (km s$^{-1}$) & (km s$^{-1}$) & (km s$^{-1}$) & candidature & (km s$^{-1}$) & run & \\
\noalign{\smallskip}
\hline
\noalign{\smallskip}
SIPS1632-0631  & -4.76$\pm$0.54 & 25.9 $\pm$ 3.2  &  14.1 $^{+  5.1 }_{-   4.6 }$    &  -33.0 $^{+  8.0 }_{-   9.0 }$    &  -27.7 $^{+  6.9 }_{-   7.8 }$    &  Pleiades  & 15.23 & 4 & 23\\
SIPS1758-6811  &  45.90$\pm$0.27 & 71.4 $\pm$ 8.7  &  -3.5 $^{+  7.0 }_{-   7.5 }$    &  -66.2 $^{+  8.7 }_{-   9.7 }$    &  -38.9 $^{+  6.5 }_{-   7.5 }$    &  Hyades  & 10.40 & 4 & 22\\
SIPS1949-7136  &  28.69$\pm$0.25 & 60.7 $\pm$ 7.4  &  -14.1 $^{+  9.0 }_{-   10.3 }$    &  -55.8 $^{+  10.1 }_{-   11.4 }$    &  -19.7 $^{+  5.1 }_{-   6.3 }$    &  Pleiades  & -  & 4 & 20\\
''             &  & 70.6 $\pm$ 8.6  &  -19.7 $^{+  10.5 }_{-   12.0 }$    &  -62.5 $^{+  11.7 }_{-   13.3 }$    &  -20.6 $^{+  5.9 }_{-   7.3 }$    &  IC 2391  & & &\\
''             &  & 46.2 $\pm$ 5.6  &  -6.0 $^{+  6.9 }_{-   7.9 }$    &  -46.0 $^{+  7.7 }_{-   8.7 }$    &  -18.5 $^{+  3.9 }_{-   4.8 }$    &  Hyades  & & &\\
SIPS2000-7523  & 11.77$\pm$0.97 & 23.7 $\pm$ 2.9  &  -7.7 $^{+  4.2 }_{-   4.7 }$    &  -10.2 $^{+  2.6 }_{-   3.0 }$    &  -21.7 $^{+  4.0 }_{-   4.5 }$    &  {\bf Castor}   & - & 4 & 22\\
2MASS2001-5949  & -19.58$\pm$0.53 & 43.9 $\pm$ 5.4  &  -36.2 $^{+  6.1 }_{-   7.0 }$    &  2.9 $^{+  5.1 }_{-   6.0 }$    &  -18.4 $^{+  6.8 }_{-   7.6 }$    &  Castor  & 19.10 & 4 & 20\\
SIPS2014-2016  & -46.29$\pm$0.45 & 33.6 $\pm$ 4.1  &  -55.2 $^{+  4.6 }_{-   5.2 }$    &  -25.9 $^{+  4.5 }_{-   5.4 }$    &  -17.7 $^{+  8.2 }_{-   9.1 }$    &  IC 2391   &  16.18 & 4 & 22\\
''           &  &  22.0 $\pm$ 2.7  &  -49.3 $^{+  3.1 }_{-   3.5 }$    &  -22.5 $^{+  3.0 }_{-   3.6 }$    &  -4.3 $^{+  5.5 }_{-   6.1 }$    &  {\bf Hyades}   & & &\\
''           &  &  25.0 $\pm$ 3.1  &  -50.8 $^{+  3.5 }_{-   3.9 }$    &  -23.4 $^{+  3.4 }_{-   4.0 }$    &  -7.8 $^{+  6.2 }_{-   6.9 }$    &  Castor  & & &\\
2MASS2031-5041 & -30.93$\pm$0.50  &  48.3 $\pm$ 5.9  &  -53.2 $^{+  8.0 }_{-   9.2 }$    &  -26.5 $^{+  9.9 }_{-   11.5 }$    &  -11.4 $^{+  8.1 }_{-   9.3 }$    &  {\bf Hyades}  & 13.00 & 4 & 19\\
SIPS2039-1126  & -18.01$\pm$2.00  & 56.0 $\pm$ 6.8  &  -13.5 $^{+  4.0 }_{-   4.1 }$    &  -31.4 $^{+  5.8 }_{-   6.3 }$    &  -14.9 $^{+  6.3 }_{-   7.0 }$    &  {\bf Pleiades}  & $>$15 & 4 & 20\\
''          &   &  48.5 $\pm$ 5.9  &  -13.4 $^{+  3.6 }_{-   3.7 }$    &  -28.4 $^{+  5.1 }_{-   5.6 }$    &  -11.8 $^{+  5.6 }_{-   6.2 }$    &  Castor  &  & &\\
SIPS2045-6332 & 0.53$\pm$0.53 &  20.0 $\pm$ 2.4  &  -11.8 $^{+  2.8 }_{-   3.1 }$    &  -17.1 $^{+  3.3 }_{-   3.5 }$    &  -4.2 $^{+  1.7 }_{-   2.0 }$    &  {\bf Pleiades} & $>$15 & 4 & 29\\
''         &    &  17.4 $\pm$ 2.1  &  -10.2 $^{+  2.5 }_{-   2.7 }$    &  -14.9 $^{+  2.9 }_{-   3.1 }$    &  -3.7 $^{+  1.6 }_{-   1.8 }$    &  {\bf Castor}  & & &\\
SIPS2049-1716 & -35.38$\pm$2.40  &  20.1 $\pm$ 2.4  &  -44.0 $^{+  6.5 }_{-   7.2 }$    &  -24.6 $^{+  3.6 }_{-   4.0 }$    &  -11.7 $^{+  8.2 }_{-   9.0 }$    &  IC 2391   & - & 1 & 17\\
''         &    &  13.2 $\pm$ 1.6  &  -37.6 $^{+  4.8 }_{-   5.3 }$    &  -21.1 $^{+  2.7 }_{-   3.0 }$    &  -0.9 $^{+  5.8 }_{-   6.4 }$    &  {\bf Hyades}  & & &\\
SIPS2049-1944  & 1.62$\pm$1.00 &  36.9 $\pm$ 4.5  &  -6.3 $^{+  2.7 }_{-   3.0 }$    &  -42.6 $^{+  6.8 }_{-   7.2 }$    &  -38.9 $^{+  6.5 }_{-   6.8 }$    &  Pleiades & 18.71  & 4 & 31\\
SIPS2100-6255  & -18.23$\pm$0.41 & 66.1 $\pm$ 10.2  &  -36.2 $^{+  7.0 }_{-   8.1 }$    &  -24.8 $^{+  9.3 }_{-   10.8 }$    &  3.4 $^{+  4.4 }_{-   5.5 }$    & {\bf Hyades}   & - & 4 &10\\
2MASS2106-4044  & -50.05$\pm$0.38 &  59.3 $\pm$ 7.2  &  -65.4 $^{+  7.7 }_{-   8.9 }$    &  -23.4 $^{+  9.4 }_{-   11.3 }$    &  1.9 $^{+  8.7 }_{-   10.0 }$    &  IC 2391  & 22.12 & 4 & 21\\
''              &  & 38.8 $\pm$ 4.7  &  -55.6 $^{+  5.2 }_{-   5.9 }$    &  -15.6 $^{+  6.2 }_{-   7.4 }$    &  12.9 $^{+  5.8 }_{-   6.6 }$    &  Hyades  & & &\\
SIPS2114-4339  & 2.69$\pm$0.25 & 35.4 $\pm$ 4.3  &  -4.2 $^{+  2.2 }_{-   2.5 }$    &  -25.0 $^{+  4.8 }_{-   5.3 }$    &  -7.1 $^{+  2.1 }_{-   2.5 }$    &  {\bf Castor}  & 25.22  & 4 & 18\\
SIPS2119-0740  & -45.50$\pm$0.34 &  61.7 $\pm$ 7.5  &  -43.7 $^{+  5.3 }_{-   6.2 }$    &  -54.8 $^{+  6.1 }_{-   6.7 }$    &  -17.8 $^{+  8.6 }_{-   9.4 }$    &  Pleiades  &  16.35 & 4 & 17\\
''           &  &  71.9 $\pm$ 8.8  &  -46.5 $^{+  6.2 }_{-   7.2 }$    &  -59.6 $^{+  7.1 }_{-   7.8 }$    &  -25.1 $^{+  10.0 }_{-   11.0 }$    &  IC 2391  & & &\\
HB2124-4228  & -7.64$\pm$0.33 & 35.7 $\pm$ 10.1  &  -16.4 $^{+  6.7 }_{-   9.3 }$    &  -29.4 $^{+  12.5 }_{-   15.7 }$    &  -4.7 $^{+  6.4 }_{-   8.9 }$    &  {\bf Pleiades}  &  17.61 & 4 & 20\\
SIPS2128-3254  & -19.77$\pm$2.15  & 33.0 $\pm$ 1.0  &  -48.0 $^{+  4.5 }_{-   4.6 }$    &  -31.1 $^{+  3.5 }_{-   3.7 }$    &  -24.2 $^{+  4.5 }_{-   4.6 }$    &  Hyades  & $\approx13$$^{1}$ & 1 & 24\\
DENIS2200-3038  & -25.00$\pm$0.14 & 37.9 $\pm$ 4.6  &  -40.0 $^{+  8.3 }_{-   9.8 }$    &  -21.0 $^{+  7.7 }_{-   9.2 }$    &  -2.5 $^{+  6.3 }_{-   7.2 }$    &  IC 2391 & 14.86   & 4 & 30\\
 ''          &                   &  24.8 $\pm$ 3.0  & -31.2 $^{+  5.5 }_{-   6.4 }$    &  -15.3 $^{+  5.0 }_{-   6.1 }$    &  5.2 $^{+  4.1 }_{-   4.8 }$    &  {\bf Hyades} &  & &\\
 ''           &                  &  28.2 $\pm$ 3.4  & -33.5 $^{+  6.2 }_{-   7.3 }$    &  -16.8 $^{+  5.7 }_{-   6.9 }$    &  3.2 $^{+  4.7 }_{-   5.4 }$    &  Castor &  & &\\
SIPS2200-2756  & -17.43$\pm$0.34 & 35.5 $\pm$ 4.4  &  -30.2 $^{+  4.4 }_{-   4.9 }$    &  -8.9 $^{+  2.5 }_{-   3.1 }$    &  -2.2 $^{+  3.4 }_{-   3.8 }$    &  {\bf Hyades} &  $<$10  & 4 & 22\\
2MASS2207-6917  & -9.21$\pm$0.41 & 56.6 $\pm$ 6.9  &  -35.7 $^{+  8.8 }_{-   10.1 }$    &  -13.1 $^{+  7.4 }_{-   8.8 }$    &  -8.1 $^{+  7.3 }_{-   8.7 }$    &  Pleiades  &  11.66 & 4 & 22\\
 ''         &    &  65.9 $\pm$ 8.0  &  -40.6 $^{+  10.2 }_{-   11.8 }$    &  -16.0 $^{+  8.6 }_{-   10.2 }$    &  -10.4 $^{+  8.4 }_{-   10.1 }$    &  IC 2391  & & &\\
''          &   &  43.2 $\pm$ 5.3  &  -28.5 $^{+  6.8 }_{-   7.8 }$    &  -9.0 $^{+  5.7 }_{-   6.7 }$    &  -4.7 $^{+  5.6 }_{-   6.7 }$    &  {\bf Hyades}  & & &\\
''          &    &  49.1 $\pm$ 6.0  &  -31.6 $^{+  7.7 }_{-   8.8 }$    &  -10.8 $^{+  6.4 }_{-   7.6 }$    &  -6.2 $^{+  6.3 }_{-   7.6 }$    &  Castor  & & &\\
LEHPM4480  & 10.43$\pm$0.30  & 51.9 $\pm$ 6.3  &  -50.0 $^{+  9.2 }_{-   9.9 }$    &  -45.6 $^{+  8.0 }_{-   8.9 }$    &  -27.5 $^{+  5.3 }_{-   6.0 }$    &  Hyades  &  14.75 & 4 &  20\\
2MASS2222-4919 & 33.76$\pm$0.38 &  100.9 $\pm$ 12.3  &  -2.3 $^{+  11.7 }_{-   14.1 }$    &  -68.0 $^{+  18.8 }_{-   21.9 }$    &  -30.6 $^{+  9.6 }_{-   10.4 }$    &  Pleiades  &  $<$10 & 4 & 20\\
2MASS2231-4443  & -7.86$\pm$0.24 & 70.7 $\pm$ 8.6  &  -49.1 $^{+  12.4 }_{-   14.3 }$    &  -14.6 $^{+  9.6 }_{-   11.8 }$    &  -20.8 $^{+  8.4 }_{-   9.7 }$    &  IC 2391  &  $<$10 & 4 & 24\\
''              & &  46.3 $\pm$ 5.6  &  -33.6 $^{+  8.2 }_{-   9.4 }$    &  -9.3 $^{+  6.3 }_{-   7.7 }$    &  -11.3 $^{+  5.6 }_{-   6.4 }$    &  {\bf Hyades}  & & &\\
''              & & 52.6 $\pm$ 6.4  &  -37.6 $^{+  9.3 }_{-   10.7 }$    &  -10.7 $^{+  7.2 }_{-   8.8 }$    &  -13.8 $^{+  6.3 }_{-   7.3 }$    &  Castor &  & &\\
LEHPM4908  & -4.15$\pm$0.22 & 49.4 $\pm$ 6.0  &  -46.4 $^{+  7.6 }_{-   8.1 }$    &  -22.8 $^{+  5.3 }_{-   5.8 }$    &  -14.7 $^{+  4.4 }_{-   5.0 }$    &  IC 2391  & - & 4 & 21\\
''         &  & 32.3 $\pm$ 3.9  &  -31.1 $^{+  5.0 }_{-   5.4 }$    &  -14.2 $^{+  3.5 }_{-   3.9 }$    &  -8.6 $^{+  2.9 }_{-   3.3 }$    &  {\bf Hyades}  & & &\\
2MASS2242-2659  & 2.97$\pm$0.21 &  72.4 $\pm$ 8.8  &  -24.5 $^{+  8.1 }_{-   9.5 }$    &  -16.5 $^{+  7.4 }_{-   8.9 }$    &  -19.3 $^{+  4.4 }_{-   5.1 }$    &  IC 2391 &  22.14  & 4& 20\\
''              & & 47.4 $\pm$ 5.8  &  -15.6 $^{+  5.3 }_{-   6.2 }$    &  -10.6 $^{+  4.9 }_{-   5.8 }$    &  -13.5 $^{+  3.0 }_{-   3.4 }$    &  {\bf Hyades}  & & &\\
''             &  & 54.0 $\pm$ 6.6  &  -17.9 $^{+  6.1 }_{-   7.1 }$    &  -12.1 $^{+  5.5 }_{-   6.6 }$    &  -15.0 $^{+  3.4 }_{-   3.8 }$    &  {\bf Castor}  & & &\\
2MASS2254-3228 & 0.15$\pm$0.55  &  65.2 $\pm$ 7.9  &  -6.5 $^{+  6.3 }_{-   7.8 }$    &  -29.6 $^{+  9.1 }_{-   10.7 }$    &  -6.6 $^{+  3.4 }_{-   4.0 }$    &  {\bf Pleiades}  &  18.62  & 4 & 22\\
''            &  &  56.5 $\pm$ 6.9  &  -5.7 $^{+  5.5 }_{-   6.8 }$    &  -25.7 $^{+  7.9 }_{-   9.3 }$    &  -5.8 $^{+  3.1 }_{-   3.6 }$    &  {\bf Castor}  & & & \\
2MASS2311-5256 & 3.33$\pm$0.11  & 87.3 $\pm$  10.6 & -45.6$^{+  13.1}_{-   15.1}$  &      -36.3$^{+   14.1}_{-   16.8}$  &     -18.2$^{+    7.5}_{-    9.1}$ & Pleiades & $<$10 & 2 & 20\\
''  & & 101.6 $\pm$ 12.4 & -53.3$^{+   15.3}_{-   17.6}$  &      -42.2$^{+   16.4}_{-   19.6}$  &     -20.7$^{+    8.8}_{-   10.6}$   & IC 2391 & & & \\
''  & & 66.5 $\pm$ 8.1 & -34.4$^{+   10.0}_{-   11.6}$  &      -27.9$^{+   10.8}_{-   12.8}$  &     -14.5$^{+    5.8}_{-    6.9}$   & {\bf Hyades} & & & \\
SIPS2318-4919   & -11.61$\pm$0.40 & 48.9 $\pm$ 3.7 & -49.4$^{+    7.3}_{-    7.9}$  &      -21.4$^{+    6.1}_{-    6.8}$  &      -6.3$^{+    3.9}_{-    4.2}$ & {\bf Hyades} & 16.50  & 2 & 22\\
SIPS2321-6106   & 10.31$\pm$0.10 & 64.9 $\pm$ 7.9 & -27.4$^{+    6.5}_{-    7.1}$  &       -0.8$^{+    3.8}_{-    3.1}$  &     -28.6$^{+    4.4}_{-    5.0}$  & Pleiades & $<$10 & 2 & 20\\
''  &  & 75.6 $\pm$ 9.2 & -32.7$^{+  7.5}_{-    8.3}$ &      -0.3$^{+  4.4}_{-    3.6}$ &    -32.0$^{+  5.2}_{-    5.8}$   & IC3291 & & & \\
SIPS2322-6357  & -8.00$\pm$0.26 & 74.0 $\pm$ 9.0 & -39.8$^{+   8.4}_{-    9.5}$ &     -20.1$^{+   7.6}_{-    8.9}$ &      -3.0$^{+   4.8}_{-    5.7}$ & Pleiades    & 14.77  & 2 &18\\
''  & & 86.2 $\pm$ 10.5 & -45.7$^{+    9.8}_{-   11.1}$  &      -23.9$^{+    8.9}_{-   10.4}$ &      -4.5$^{+    5.5}_{-   6.6}$ & IC 2391 & & & \\
'' & & 56.4 $\pm$ 6.9 & -31.2$^{+    6.5}_{-    7.3}$  &      -14.5$^{+    5.8}_{-    6.8}$ &      -0.8$^{+    3.7}_{-    4.4}$  & {\bf Hyades} & & &\\
SIPS2341-3550 & -2.76$\pm$0.21 & 56.0 $\pm$ 6.8 & -33.7$^{+    8.4}_{-    9.6}$    &      -23.3$^{+    7.4}_{-    8.7}$ &      -7.3$^{+    2.8}_{-    3.1}$ &  {\bf IC3291} &14.22  & 2  & 23\\
''  & & 36.7 $\pm$ 4.5 & -22.4$^{+    5.5}_{-    6.3}$  &      -15.2$^{+    4.8}_{-    5.7}$ &      -3.9$^{+  1.9}_{-    2.1}$ & Hyades&  & & \\
'' & & 41.7 $\pm$ 5.1 &-25.3$^{+    6.3}_{-    7.2}$  &      -17.3$^{+    5.5}_{-    6.5}$ &      -4.8$^{+  2.1}_{-    2.4}$ & Castor & & & \\
SIPS2343-2947   &  -19.83$\pm$0.57 & 49.3 $\pm$ 6.0 &  -56.7$^{+   10.6}_{-   11.7}$  &      -27.9$^{+    7.5}_{-   8.6}$   &     3.6$^{+  3.4}_{-   3.6}$ & IC 2391  & 19.42 & 2 & 20\\
''  & & 32.3 $\pm$  3.9 & -38.8$^{+    7.0}_{-    7.7}$  &      -18.8$^{+  4.9}_{-   5.7}$ &       9.0$^{+    2.4}_{-   2.6}$ & Hyades &  & & \\
\noalign{\smallskip}
\hline
\noalign{\smallskip}
\end{tabular}
\end{center}
\end{flushleft}
\end{table*}

\begin{table*}
\contcaption{
\label{tab:vrs3}}
\begin{flushleft}
\scriptsize
\begin{center}
\begin{tabular}{lccrrrcccc}
\noalign{\smallskip}
\hline
\noalign{\smallskip}
Name        &  $V_{\rm r}$ $\pm$ $\sigma_{V_{\rm r}}$ & distance & $U \pm \sigma_{\rm U}$  &  $V \pm \sigma_{\rm V}$ &  $W \pm \sigma_{\rm W}$              &  MG & $V\sin{i}$ & Observing & S/N \\
 & (km s$^{-1}$) & (pc) & (km s$^{-1}$) & (km s$^{-1}$) & (km s$^{-1}$) & candidature & (km s$^{-1}$) & run & \\
\noalign{\smallskip}
\hline
\noalign{\smallskip}
SIPS2347-1821   & -6.24$\pm$0.32 &  31.8 $\pm$ 3.9 & -32.7$^{+    6.0}_{-    6.6}$  &      -10.0$^{+    3.2}_{-   3.7}$ &      -1.2$^{+    1.8}_{-    2.0}$ &  {\bf Hyades} & 19.19 & 2 & 21\\
SIPS2350-6915   & -6.81$\pm$0.48 & 73.6 $\pm$ 9.0 & -53.8$^{+   12.4}_{-   14.1}$  &      -17.8$^{+    9.4}_{-  11.3}$ &     -11.0$^{+    7.2}_{-    8.6}$ & Pleiades & 21.20 & 2 & 22\\
''  & & 85.6 $\pm$ 10.4  & -62.1$^{+   14.4}_{-   16.3}$  &      -21.3$^{+   10.9}_{-   13.1}$ &     -13.6$^{+    8.4}_{-   10.0}$ & IC 2391 & & &\\
''  & & 56.1 $\pm$ 6.8 & -41.7$^{+    9.5}_{-   10.8}$  &      -12.7$^{+    7.2}_{-    8.6}$ &      -7.2$^{+    5.6}_{-    6.6}$ & {\bf Hyades} & & &  \\
 LEHPM6375 &   3.63$\pm$0.15 & 37.5 $\pm$ 4.6 & -19.7$^{+     7.5}_{-   8.9}$    &  -43.9$^{+    9.8}_{-  11.0}$   &   -15.1$^{+     2.8}_{-   3.1}$   &    Sirius &  $<$10  & 2 & 25\\
'' & & 43.3 $\pm$ 5.3  &   -22.9$^{+     8.6}_{-  10.2}$    &  -50.7$^{+    11.3}_{-  12.7}$    &  -16.9$^{+  3.2}_{-   3.5}$  &  Pleiades   & & & \\
'' & & 50.4 $\pm$ 6.2  &   -26.7$^{+    10.0}_{-  11.9}$    &  -59.1$^{+    13.1}_{-  14.7}$    &  -19.1$^{+  3.6}_{-   4.1}$  &    IC 2391  & & & \\
'' & & 33.0 $\pm$ 4.0 &    -17.3$^{+     6.6}_{-   7.8}$   &   -38.5$^{+     8.6}_{-   9.7}$   &   -13.7$^{+  2.4}_{-   2.7}$  &   {\bf  Hyades}   & & &\\
'' & & 37.5 $\pm$ 4.6  &   -19.7$^{+     7.5}_{-   8.9}$    &  -43.9$^{+     9.8}_{-  11.0}$    &  -15.1$^{+  2.8}_{-   3.1}$   &    Castor   & & &\\
LEHPM6542   & 4.02$\pm$0.71 &  64.7 $\pm$ 7.9 & -48.5$^{+   13.7}_{-   15.8}$  &      -25.2$^{+   11.5 13.7}$ &     -15.7$^{+    4.1  4.6}$  & IC 2391 & 15.61  & 2 & 23\\
'' &  & 42.3 $\pm$ 5.2 & -31.6$^{+    9.0}_{-   10.4}$  &      -16.2$^{+    7.6}_{-    9.0}$ &     -11.7$^{+    2.9}_{-    3.3}$ & {\bf Hyades} & & &\\
''  &  & 48.2 $\pm$ 5.9 &-36.0$^{+  10.2}_{-  11.8}$  &      -18.5$^{+    8.6}_{-   10.3}$ &     -12.7$^{+    3.2}_{-    3.6}$ & Castor & & & \\
\noalign{\smallskip}
\hline
\noalign{\smallskip}
\end{tabular}
\end{center}
$^{1}$ Measused with low S/N\\
 '*' See Sect. 5.3\\ 
\end{flushleft}
\end{table*}

\begin{table*}
\caption[Results]{Results: Firts and second columns are name and kinematic classification. 
 Third column classifies the objects as young (Y) if they lie in the upper part of the limiting line
 in rotational velocity, old (O) if they lie under the line and unknown age (Y-O) if they
 are in the limit of both areas or the rotational velocity superior limit
 is not clear enough to classify them in one of the groups. 
 Column four highlights to which MG the candidates are kinematic members. 
 HY for Hyades, SI for Sirius, CA for Castor, PL for Pleiades and
 IC for IC 2391, and OYD for other young disk members without a membership
 of one of the five MG in consideration. The '?' marks object where a further
 spectroscopic study is need to asses or dismiss its membership. (N) markes 
 the two objects that have been dismissed due to the rotational velocity criterion.
\label{tab:criteria}}
\begin{flushleft}
\scriptsize
\begin{center}
\begin{tabular}{lccc}
\hline
\noalign{\smallskip}
Name & Kinematic &  $V\sin{i}$ Classification  & Possible MG membership \\
\noalign{\smallskip}
\hline
\noalign{\smallskip}
SIPS0004-5721   &  YD & Y &  {\bf  HY, IC } \\
SIPS0007-2458   &  YD & Y &  {\bf  IC }\\
2MASS0020-2346   &  YD & Y &  {\bf OYD} \\
DENIS0021-4244   &  YD & O &  (N)  \\
SIPS0027-5401   &  YD & Y & {\bf  HY } \\
SIPS0039-2256   &  YD & Y-O & {\bf OYD?}  \\
DENIS0041-5621   & YD  & Y  & {\bf  OYD}\\
SIPS0054-4142   &  YD & Y-O  & {\bf OYD?} \\
SIPS0109-0343   &  YD  & O  & - \\
LEHPM1289      &  OD  & Y  &  - \\
SIPS0115-2715   &  OD &  Y-O & -  \\
2MASS0123-3610*   & YD  & Y & {\bf  HY } \\
SIPS0126-1946   &  OD & Y & - \\
LEHPM1563      &  YD & Y  &  {\bf  OYD} \\
SIPS0153-5122   &  YD & Y & {\bf   HY, IC } \\
2MASS0204-3945   &  YD & O &  (N)  \\
SIPS0212-6049   &  OD & Y-O &  -\\
SIPS0214-3237   &  YD & Y  & {\bf  HY}\\
SIPS0235-0711   &   YD &  Y  & {\bf  HY}\\
2MASS0334-2130   &  YD & Y-O  &  {\bf IC?}  \\
2MASS0429-3123   &  YD & O  & - \\
SIPS0440-0530   &  YD &  Y &  {\bf  HY}\\
2MASS0445-5321   & OD &  - &  - \\
2MASS0502-3227   & OD  & Y-O  & - \\
2MASS0528-5919   & OD  & Y-O  & - \\
2MASS0600-3314   & YD &  - &  {\bf HY? }\\
SIPS1039-4110   &  YD & -  &  {\bf OYD? }\\
SIPS1124-2019   &  YD & -  & - \\
DENIS1250-2121   &  OD & -  & -\\
SIPS1329-4147   & OD &  - &  - \\
SIPS1341-3052   &  OD &  - & - \\
2MASS1507-2000   & OD &  - & -  \\
SIPS1632-0631   &  OD & O  &  -\\
SIPS1758-6811   &  OD & Y  & - \\
SIPS1949-7136   &  OD & -  & - \\
SIPS2000-7523   & YD & -  &  {\bf  CA? }\\
2MASS2001-5949   & OD  & Y  & - \\
SIPS2014-2016   & YD  & Y  & {\bf  HY }\\
2MASS2031-5041   &   YD &  Y  & {\bf  HY } \\
SIPS2039-1126   & YD  & Y  &  {\bf  PL }\\
SIPS2045-6332   & YD  & Y-O  &  {\bf  CA, PL? }\\
SIPS2049-1716   & YD  & -  & {\bf  HY?  }\\
SIPS2049-1944   & YD  & Y  & {\bf  OYD}\\
SIPS2100-6255    & YD    & -   & {\bf  HY? }\\
2MASS2106-4044   & OD  & Y  & - \\
SIPS2114-4339   &  YD & Y  & {\bf  CA }\\
SIPS2119-0740   & OD  & Y  &  - \\
HB2124-4228   & YD  & Y  & {\bf  PL } \\
SIPS2128-3254   &   YD & Y  & {\bf  OYD}\\
DENIS2200-3038   &  YD &  O & HY? \\
SIPS2200-2756   & YD  &  Y-O & {\bf  HY?  } \\
2MASS2207-6917   & YD  &  Y & {\bf  HY }\\
LEHPM4480   & OD  & Y  & -  \\
2MASS2222-4919   &  OD & O & - \\
2MASS2231-4443   &  YD & Y & {\bf  HY } \\
LEHPM4908   & YD  & -  &  {\bf  HY?  } \\
2MASS2242-2659   &  YD &  Y  &  {\bf  CA, HY } \\
2MASS2254-3228   & YD  &  Y  & {\bf  CA, PL } \\
2MASS2311-5256   &  YD &  Y  & {\bf  HY } \\
SIPS2318-4919   & YD & Y  & {\bf  HY } \\
SIPS2321-6106   & YD  & Y &  {\bf  OYD} \\
SIPS2322-6357   & YD  & Y  & {\bf  HY } \\
SIPS2341-3550   & YD  & Y  & {\bf  IC } \\
SIPS2343-2947   &  YD  & Y  &  {\bf  OYD}  \\
SIPS2347-1821   & YD  &  Y  & {\bf  HY } \\
SIPS2350-6915   &  YD & Y  & {\bf  HY } \\
LEHPM6375   &  YD &  Y-O  & {\bf  HY?  } \\
LEHPM6542   & YD &  Y  & {\bf  HY } \\
\noalign{\smallskip}
\hline
\end{tabular}
\end{center}
'*' See Sect. 5.3 \\
\end{flushleft}
\end{table*}

\section*{Acknowledgments}
 M.C. G\'alvez-Ortiz acknowledges financial support from the European Commission
 in the form of a Marie Curie Intra European Fellowship (PIEF-GA-2008-220679).
 Based on observations made with ESO Telescopes, with FEROS echelle spectrometer mounted on        
 the 2.2-m telescope based at La Silla (Chile) under programmes 078.C-0333(A) and 079.C-0255(A),
 and with UVES high-resolution optical spectrograph at VLT Kueyen 8.2-m telescope                  
 at Paranal (chile) in programmes 079.C-0308(A), 080.C-0923(A) and 081.C-0222(A).
 And Observations obtained at the Gemini Observatory, which is operated by the
Association of Universities for Research in Astronomy, Inc., under a cooperative agreement
with the NSF on behalf of the Gemini partnership: the National Science Foundation (United
States), the Science and Technology Facilities Council (United Kingdom), the
National Research Council (Canada), CONICYT (Chile), the Australian Research Council
(Australia), Ministério da Ciência e Tecnologia (Brazil)
and Ministerio de Ciencia, Tecnología e Innovación Productiva (Argentina),
 with the Phoenix infrared spectrograph, developed and operated by the National Optical Astronomy
Observatory in program GS-2008A-C-1.

\bsp

\label{lastpage}

\end{document}